\documentclass[nofootinbib,aps,prd,preprint,superscriptaddress]{revtex4-1}

\usepackage{amsmath}
\usepackage{epsfig}
\usepackage{graphicx}
\usepackage{color}
\usepackage{latexsym}
\usepackage{amssymb}

\def\nn{\nonumber}
\def\TeV{\,{\rm TeV}}
\def\GeV{\,{\rm GeV}}
\def\MeV{\,{\rm MeV}}

\newcommand{\n}{\overline{n}}
\newcommand{\hs}{\hat{s}}

\newcommand{\mO}{\mathcal{O}}

\newcommand{\mB}{\mathcal{B}}
\newcommand{\mL}{\mathcal{L}}

\newcommand{\mW}{\mathcal{W}}
\newcommand{\mY}{\mathcal{Y}}

\newcommand{\be}{\begin{equation}}
\newcommand{\ee}{\end{equation}}
\newcommand{\bea}{\begin{eqnarray}}
\newcommand{\eea}{\end{eqnarray}}
\newcommand{\ti}{\tilde{t}}
\newcommand{\tiv}{\tilde{t}_v}
\newcommand{\timv}{\tilde{t}_{-v}}
\newcommand{\mt}{m_{\ti}}

\begin{document}

\baselineskip 3.0ex
\vspace*{18pt}

\begin{flushright}
{\small KIAS-P14011}
\end{flushright}


\title{Production of Stoponium at the LHC}

\def\Seoultech{Institute of Convergence Fundamental Studies and School of Liberal Arts, Seoul National University of Science and Technology, Seoul 139-743, Korea}
\def\Duke{Department of Physics, Duke University, Durham NC 27708, USA}
\def\KIAS{School of Physics, KIAS, Seoul 130-722, Korea}

\author{Chul Kim}
\email[E-mail:]{chul@seoultech.ac.kr}
\affiliation{\Seoultech}
\author{Ahmad Idilbi}
\email[E-mail:]{aui13@psu.edu}
\affiliation{ Physics Department , Pennsylvania State University, University Park, PA, 16802, USA}
\author{Thomas Mehen}
\email[E-mail:]{mehen@phy.duke.edu}
\affiliation{\Duke}
\author{Yeo Woong Yoon}
\email[E-mail:]{ywyoon@kias.re.kr}
\affiliation{\KIAS\vspace{0.5cm}}

\begin{abstract}
\baselineskip 3.0ex  \vspace{0.5cm}
Although the Large Hadron Collider (LHC) has not observed supersymmetric (SUSY) partners of the Standard Model particles, their existence is not  ruled out yet.
One recently explored scenario in which there are light SUSY partners that have evaded current bounds from the LHC is that of a light long-lived stop quark.
In this paper we consider light stop pair production at the LHC when the stop mass is between 200 and $400\GeV$. If the stops are long-lived they can form a bound state, stoponium, which then undergoes two-body decays to Standard Model particles. By considering the near-threshold production of such a pair through the gluon-gluon fusion process and taking into account the strong Coulombic interactions responsible for the formation of this bound state, we obtain factorization theorems for the stop pair inclusive  and differential production cross sections. We also perform a resummation of large  threshold logarithms up to next-to-next-to-leading logarithmic accuracy using well-established  renormalization group equations in an effective field theory methodology. These results are used to calculate the invariant mass distributions  of two photons or two $Z$ bosons coming from the decay of  the stoponium at the LHC. For our choices of SUSY model parameters, the stoponium is not detectable above Standard Model backgrounds in $\gamma \gamma$ or $ZZ$ at $8\TeV$, but will be visible with $400\,{\rm fb}^{-1}$ of accumulated data if its  mass is below $500\GeV$ when the LHC runs at $14\TeV$.
\end{abstract}

\maketitle


\section{Introduction}
The  Large Hadron Collider (LHC) has confirmed the existence of the Higgs boson. The next main objective of future searches is to discover what lies beyond the Standard Model (BSM).
Supersymmetric (SUSY) extensions of the SM are among the most well-studied scenarios for BSM physics, and are  mainly motivated by
 theoretical considerations (e.g., the hierarchy problem). Although the LHC has not found any signals of BSM physics so far,  SUSY extensions of the SM are not ruled out yet.
Among a variety of SUSY scenarios, the possibility of a light stop has received a lot of attention since it is naturally well motivated~\cite{Dimopoulos:1995mi,Pomarol:1995xc,Cohen:1996vb,Kitano:2006gv,Kats:2011qh,Brust:2011tb,Brust:2012uf,Papucci:2011wy,Espinosa:2012in}. A light stop is generally expected in the minimal supersymmetric Standard Model (MSSM) due to a large positive Yukawa coupling term in the  renormalization group equation for the stop mass and possibly large mixing term between   left and right stops in the
stop mass matrix. On top of that, there are other motivations for the light stop coming from  cosmological considerations. First, a light stop with a mass that is a few tens of $\GeV$ above the lightest SUSY particle can successfully account for the thermal relic density of dark matter \cite{Boehm:1999bj}. Second, electroweak baryogenesis is possible in the MSSM with a light stop \cite{Huet:1995sh,Carena:2008vj,Li:2008ez}. Recently, the authors of Ref. \cite{Delgado:2012eu}  have found that a light stop mass between 200 and $400\GeV$ (with several additional conditions) is consistent with the currently available experiment constraints such as the $126\GeV$ Higgs mass, $B\to X_s \gamma$, etc.

Collider experiments search for stop quarks decaying to a top quark and neutralino  or bottom quark and chargino and
if the stop is light enough that these decays are not kinematically
allowed, then the bounds from these searches are not relevant and the stop quark is likely long-lived.
See Refs.~\cite{Cao:2012rz,Kribs:2013lua,Kowalska:2013ica,Han:2013kga} for light stop searches from stop decays to top quark and neutralino or bottom quark and chargino.
This SUSY spectrum can be obtained if the
bino and wino masses are comparable to the Higgsino mass term so that the lightest neutralino
and chargino masses are not degenerate.
 If  stops  exist, they will be produced at the LHC mainly through gluon-gluon fusion, much like the main production channel of the SM
 Higgs boson. The stops can either be produced singly or in pairs.
  If pair produced, they can  then form a bound state, stoponium, through  the strong Coulomb interaction.
 This bound state can then undergo two-body decays to SM particles, and the bound state will appear as a resonance above the SM background  for $\gamma \gamma$, $W^+W^-$, or $Z^0 Z^0$, for example. In earlier work~\cite{Kim:2008bx,Idilbi:2010rs}, we showed that these are good channels in which to search for heavy BSM particles that are strongly interacting. In this work we apply the methodology of Refs.~\cite{Kim:2008bx,Idilbi:2010rs} to stop pairs. For early work advocating searching for stoponium in the $\gamma \gamma$ channel,
 see Refs.~\cite{Drees:1993uw,Martin:2008sv}.

Since the LHC is a hadronic machine where two protons collide at very high energies, the partons inside the hadrons will initiate a hard reaction
responsible for the production of massive particles. To separate nonperturbative  long-distance QCD effects  from the calculable  short-distance effects, we derive factorization theorems for the production process. These  theorems clarify  what is perturbatively   calculable
and what is the proper form of the relevant hadronic matrix elements to be determined from experiment (or a  nonperturbative QCD calculation). In cases where such factorization theorems  hold, one then needs to deal with large logarithms encountered in perturbative calculations. Such large logarithms exist because  the production process is characterized by several  widely separated scales  and thus large logarithms of the ratios of these scales need to be resummed.

In this work we utilize the effective field theory approach to establish a factorization theorem for the production of massive stop pairs. This
is done by constructing  effective operators, at each relevant scale, that mediate the production reaction.   In our work, the relevant  theories are soft-collinear effective theory (SCET) \cite{SCET1,SCETf,Bauer:2003mga}
and heavy-scalar effective theory (HSET).
The former  describes the multiscale physics behind the production of the stop pair through the gluon-gluon fusion process. The factorization of the production process into hard, soft, and collinear parts allows us to implement the threshold resummation by solving the renormalization group equations for each of these parts. HSET describes the production of a slowly moving stop pair whose strong Coulomb interactions will bind the stop pair into the stoponium.
The strong Coulomb interactions are resummed to all orders using the Coulomb Green's function and including the finite width of the stoponium yields a resonant shape in the vicinity of the stoponium mass~\cite{Beneke:2010da,Falgari:2012hx}. For recent next-to-next-to-leading logarithmic (NNLL) resummed calculations of squark and gluino production, including the Coulomb Green's function, see Ref.~\cite{Beneke:2013opa}.
Threshold resummation of squark and gluino production in Mellin space has been calculated up to NNLL accuracy in Refs.~\cite{Langenfeld:2009eg,Langenfeld:2010vu,Langenfeld:2012ti,Pfoh:2013iia,Beenakker:2010nq,Beenakker:2011sf,Beenakker:2013mva}.  NNLL resummation in momentum space for stop pair production was recently reported in Refs.~\cite{Broggio:2013uba,Broggio:2013cia}.

The factorization theorem obtained allows us to resum large logarithms when the stop pair is produced near the partonic threshold which is the
 kinematical range of interest once we assume a light stop mass and LHC energies.
The accuracy of resummation depends on the knowledge we have of the perturbatively calculable anomalous dimensions and beta functions appearing in the formulas
 for the resummed cross section. In this work the resummation is performed up to NNLL accuracy. The phenomenological impact of resummation is discussed below.

Finally, we consider the decay rates of the stoponium bound state, denoted $\tilde \sigma$, in two channels: $pp\to\tilde{\sigma}\to\gamma\gamma$ and $pp\to\tilde{\sigma}\to ZZ$
which, as we will argue below, are the most promising channels for searching for  stoponium. Current bounds on the stop mass are frequently presented as exclusion plots in the neutralino mass - stop mass plane, with stop masses being excluded up to 700 GeV for certain neutralino masses. A gap in the exclusion plots exists wherever the stop mass is less than (approximately)  the sum of the top mass and neutralino mass, and this gap extends down to stop masses of about 200 GeV. (The plots we are referring
to can be seen in Ref.~\cite{ATLAS:directStop}. See also  Ref.~\cite{Chatrchyan:2013xna} for recent direct stop search at CMS.) For this reason, in this paper we focus on light stop masses lying between 200 and $400\GeV$.
We also study the dependence of the resonant cross section on MSSM parameter choice while taking into account uncertainties resulting from different choices of the hard, soft, and factorization scales. As expected NNLL resummation  greatly reduces these scale uncertainties.

We also discuss the required luminosity at LHC energies, 8 and $14\TeV$, and consider five different stoponium masses ranging from 400 to $800\GeV$.
Our findings show that, independent of the MSSM parameter space, we can determine whether  stoponium resonance of mass up to $500\GeV$ exists or not within the first LHC run
at  $14\TeV$, assuming  400 fb$^{-1}$ of integrated luminosity,  through either the $\gamma\gamma$ or $ZZ$ decay modes.
Based on our analysis we could not exclude any stoponium mass within that mass range with the currently accumulated LHC data at $8\TeV$.

This paper is organized as follows. In Sec.~II, we outline the theoretical framework for deriving the effective Lagrangian for the production of massive stops through the strong interactions. In Sec.~III, we consider the near-threshold  production cross section for $\tilde {\sigma}$ in $pp$ collisions and obtain the factorization theorem for this process.  This includes the   Green's function responsible for resumming the strong Coulomb interaction. In Sec.~IV, we derive the
 cross section including threshold resummation for
$pp \rightarrow \tilde{\sigma}X$ followed by  two-body decays to SM particles. This resummation is performed directly  in momentum space. In Sec.~V, we present our phenomenological results. These  include plots of  the  branching fractions for stoponium to various two-body SM final states  and
 cross sections for $pp \to \tilde \sigma \to \gamma \gamma, ZZ$. We study these as functions of the stoponium mass,  for various choices of MSSM parameters, and for two different
 LHC collision energies, 8 and 14 TeV. We conclude in Sec.~VI.
In Appendix A, we give the expressions for the  rates for stoponium decaying  into two SM particles.
 In Appendix B we collect all the formulas for the anomalous dimensions and beta functions needed to obtain the NNLL threshold resummation for our cross section.
Appendix C contains the explicit formulas for the next-to-leading order (NLO) Coulomb Green's function.

\section{Effective Lagrangian Near Threshold}

At the LHC, stop pair is produced dominantly via the $gg$ fusion process through strong interactions.
The gluons couple to the stops via the kinetic terms for the stops in the MSSM Lagrangian,
\be
\label{LQCD}
\mL_{\mathrm{\tilde t}} = - \ti^{\dagger} D^2 \ti - m_{\ti}^2 ~\ti^{\dagger} \ti,
\ee
where
 $D_{\mu} = \partial_{\mu} - ig A_{\mu}$
and $\ti$ is the scalar top (stop) field. Near threshold where the partonic center-of-mass (CM) energy
 $\hs$  is approximately $(2m_{\ti})^2$, the produced stop pair moves slowly, hence the $\tilde {t}$ can be represented
 by a nonrelativistic  heavy scalar field, analogous to the heavy quark field in heavy quark effective theory (HQET), where the velocity becomes a label of the quantum field.
 The fields of this Heavy Scalar Effective Theory (HSET) are related to the full theory fields by
\bea\label{sqmat}
\ti(x) &=&\frac{1}{\sqrt{2\mt}} (e^{-i\mt v\cdot x} \tiv (x) + e^{i\mt v\cdot x} \timv (x)), \\
\ti^{\dagger} (x) &=&\frac{1}{\sqrt{2\mt}} (e^{i\mt v\cdot x} \tiv^{\dagger} (x) + e^{-i\mt v\cdot x} \timv^{\dagger} (x)). \nonumber
\eea
The $\tiv$ and $\timv$ are the HSET fields for stop and antistop respectively.

Putting Eq.~(\ref{sqmat}) into Eq.~(\ref{LQCD}), and dropping terms with nontrivial exponentials  that vanish in the $m_{\tilde t} \to \infty$ limit,  we obtain the HSET Lagrangian up to $O(1/\mt)$:
\bea\label{HSET}
\mL_{\mathrm{HSET}} &=& \tiv^{\dagger} v\cdot iD \tiv - \frac{1}{2\mt} \tiv^{\dagger} D^2 \tiv \\
&&+\timv^\dagger(-v)\cdot iD \timv - \frac{1}{2\mt} \timv^{\dagger} D^2 \timv.  \nonumber
\eea
Here the second and fourth terms are suppressed by $\mO(1/m_{\tilde t})$. The covariant derivative in Eq.~(\ref{HSET}) can be written as
\be
\label{devsp}
D^{\mu} = \partial_s^{\mu} - igA_s^{\mu} + \partial_p^{\mu}- igA_p^{\mu} = D_s^{\mu}+D_p^{\mu},
\ee
where $A_s$ is the soft gluon, and $A_p$ is the potential gluon, the exchange of which gives rise to Coulombic potential  between the stop and antistop. We also separate the
derivatives, i.e., $\partial = \partial_s + \partial_p$ requiring $[\partial_s,A_p] = [\partial_p,A_s]=0$.

The HSET Lagrangian in Eq.~(\ref{HSET}) encodes the interaction of the stop field $\tilde{t}$ with soft and Coulomb gluons.   Those two interactions can be
 decoupled via gluon field redefinitions  where one defines hatted fields through
\begin{equation}
\label{reds}
gA_s^{\mu} = g\hat{A}_s^{\mu} + \hat{Y}_v[iD_p^{\mu}, \hat{Y}_v^{\dagger}].
\end{equation}
The hatted field is a newly defined soft field, and $\hat{Y}_v$ is the timelike soft Wilson line, which is given by
\be\label{redY}
\hat{Y}_v(x) = \mathrm{P} \exp \Bigl(ig \int^x_{-\infty} ds ~v\cdot \hat{A}_s (v s) \Bigr),
\ee
where ``P'' represents a path-ordered integral. The covariant derivative $D^{\mu}$ in Eq.~(\ref{devsp}) can be expressed in terms of $\hat{A}_s$ as
 $D^{\mu} = \hat{D}_s^{\mu} + \hat{Y}_v D_p^{\mu} \hat{Y}_v^{\dagger}$. Next we redefine the heavy stop field
as $\tiv = \hat{Y}_v \tiv^{(0)}$, where the newly defined field $\tiv^{(0)}$  does not interact with soft fields (at LO in $1/m_{\tilde t}$)
and the soft interactions of
$\tilde t$ are taken care of  by the soft Wilson line $\hat{Y}_v$. When the HSET Lagrangian is expressed in terms of the redefined fields
 $\hat{A}_s$ and $\tiv^{(0)}$ it then becomes an effective Lagrangian describing a nonrelativistic stop strongly interacting only with Coulomb  gluons.
 For convenience of notation, in the rest of the paper  we will drop the hats on these fields.

Now we construct the effective interaction Lagrangian for $gg \to \ti^{\dagger} \ti$ to be denoted below by $\mL_{\mathrm{EFT}}$. Near partonic threshold,
 only soft and collinear  gluons can be emitted into the final state. By collinear we mean collinear to one of the incoming beams. It is useful then to construct an effective operator basis in
the irreducible color representation since, in this basis, the effective operators do not mix. Since the possible irreducible
color representations of stop pair are only $\bf 1$ and $\bf 8$, the production channels allowed by color  conservation are:
 $(R_i,R_f) =\bf(1,1),~(8_S,8),~(8_A,8)$, where $R_{i}$ and $R_f$ denote the color representations of the initial and final states.
 The effective Lagrangian is then
\bea\label{EFTL}
\mL_{\mathrm{EFT}} = \sum_{k=1}^3 C_k (Q^2,\mu) \mO_k (\mu),
\eea
where $Q^2 \sim 4m_{\ti}^2$ is the typical hard scale (squared) for stop pair production, and
the effective operators $\mO_k$ are
\be\label{efto}
\mO_k = \frac{1}{2m_{\ti}^3} E_{ab\alpha\beta}^{(k)} (\mY_n \mB^{\mu}_{n\perp})^a (\mY_{\n} \mB^{\perp}_{\n\mu})^b (\tiv^{\dagger} Y_v)_{\alpha} (Y_v^{\dagger} \timv)_{\beta},
\ee
where we introduced two light-cone vectors $n$ and $\n$ for the two beam directions. They satisfy $n^2=\n^2=0$ and $n\cdot \n = 2$.
Superscripts (subscripts) $a$ and $b$ ($\alpha$ and $\beta$) are color indices in the adjoint (fundamental) representation.
 $\mB_{n\perp}^{a\mu}$ is an $n$-collinear gluon field strength tensor at LO in the SCET power counting parameter, $\lambda \sim p_{\perp}/\n\cdot p$, where $\n \cdot p$
is the large collinear momentum
component of an $n$-collinear gluon. It is defined as
\be\label{Bn}
\mB_{n\perp}^{a\mu} = i \n^{\rho} g^{\mu\nu}_{\perp} G^b_{n,\rho\nu} \mW_n^{ba}
= i\n ^{\rho} g^{\mu\nu}_{\perp} \mW_n^{\dagger,ab} G^b_{n,\rho\nu},
\ee
where $\mW_n$ is an $n$-collinear Wilson line in the adjoint representation given by
\be\label{colW}
\mW_n^{ab} (x) = \mathrm{P} \exp \Bigl(ig \int^x_{-\infty} ds ~\n\cdot A_n^c (\n s) t^c \Bigr)^{ab}\,.
\ee
The $n$-collinear gluon field is $ A_n^c$ and $(t^c)^{ab} = -if^{cab}$ is a generator in the adjoint representation.
$\mB_{\n\perp}^{a\mu}$  is defined in the same way as $\mB_{n\perp}^{a\mu}$ with $n$ and $\n$ interchanged. In Eq.~(\ref{efto}) we decoupled the soft interactions from $n$- and $\n$-collinear fields, then obtained $\mY_n$ and $\mY_{\n}$ in the adjoint representation respectively.
 These soft Wilson lines are
\bea
\mY_n^{ab} (x) &=& \mathrm{P} \exp \Bigl(ig \int^x_{-\infty} ds ~n\cdot A_s^c (n s) t^c\Bigr)^{ab}, \\
\mY_{\n}^{ab} (x) &=& \mathrm{P} \exp \Bigl(ig \int^x_{-\infty} ds ~\n\cdot A_s^c (\n s) t^c \Bigr)^{ab}.
\eea

In Eq.~(\ref{efto}) the color coefficient for each operator is defined as~\cite{Beneke:2009rj}
\be \label{ecoe}
E_{ab\alpha\beta}^{(k)}= E_{ab\alpha\beta}^{(R_i,R_f)}=\frac{C^{R_i}_{lab}{C^{R_f}_{l\alpha\beta}}^*}{\sqrt{\mathrm{dim}~R_i}},
\ee
where $C_{lab}^{R_i}$ and $C^{R_f}_{l\alpha\beta}$ are the Clebsh-Gordan coefficients for the color octet and triplet respectively, and $l$
is a dummy index running from 1 to dim$\,R_i$. The $E_{ab\alpha\beta}^{(i)}$ satisfy the orthonormality relation
\be
E_{ab\alpha\beta}^{(i)} E_{ab\alpha\beta}^{(j)\,*} = \delta^{ij}\,.
\ee
In case of $gg \to \ti^{\dagger} \ti$, the coefficients are
\bea
E^{(1)}_{ab\alpha\beta} &=& E^{\bf(1,1)}_{ab\alpha\beta}=\frac{1}{\sqrt{N_c D_A}} \delta_{ab}\delta_{\alpha\beta}, \nonumber \\
\label{coef}
E^{(2)}_{ab\alpha\beta} &=& E^{\bf(8_S,8)}_{ab\alpha\beta}=\frac{1}{\sqrt{2 B_F D_A}} D^k_{ab}T^k_{\alpha\beta}, \\
E^{(3)}_{ab\alpha\beta} &=& E^{\bf(8_A,8)}_{ab\alpha\beta}=\sqrt{\frac{2}{N_c D_A}} F^k_{ba}T^k_{\alpha\beta}, \nonumber
\eea
where $F^k_{ab} = t^k_{ab} = -if^{kab}$ is the totally antisymmetric tensor in color space and $D^k_{ab} = d^{kab}$ is the totally symmetric one.
 The color factors are  $B_F = \frac{N_c^2-4}{4N_c}$, and $D_A = N_c^2-1$.

\begin{figure}[t]
\begin{center}
\includegraphics[width=16cm]{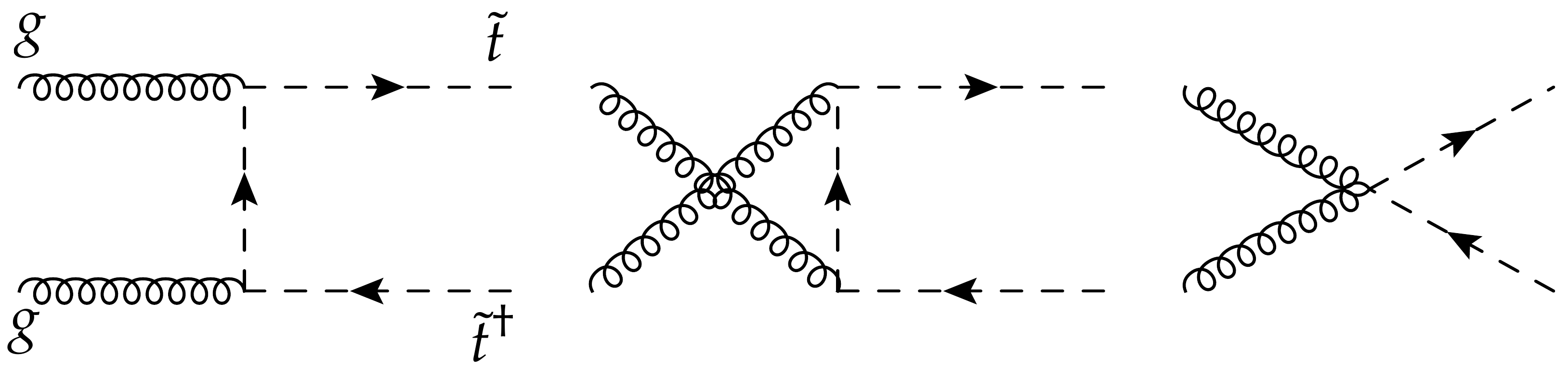}
\end{center}
\vspace{-0.3cm}
\caption{
\label{fig1}
\baselineskip 3.0ex
Tree-level processes for $gg \to \ti\ti^{\dagger}$.}
\end{figure}

We calculate the leading Wilson coefficients of operators in Eq.~$(\ref{efto})$ by computing the relevant Feynman diagrams in figure \ref{fig1}.
For transversely polarized gluons, the first two diagrams in figure \ref{fig1} are ${\cal O}(\beta^2)$, where $\beta=\sqrt{1-4m_{\tilde t}^2/{\hat s}}$, and therefore vanish at threshold.
Thus only the last diagram contributes to the matching coefficient. The results are
\bea \label{wilt}
C_1 = C_{\bf 1,1} &=& \pi\alpha_s \sqrt{\frac{D_A}{N_c}} = \sqrt{\frac{8}{3}} \pi \alpha_s \,, \nn \\
C_2 =C_{\bf 8_S,8} &=& \pi\alpha_s \sqrt{2B_F D_A} = \sqrt{\frac{20}{3}} \pi \alpha_s\,, \nn \\
C_3=C_{\bf 8_A,8}&=& 0\,.
\eea
Therefore the leading effective Lagrangian is
\be
\label{lagt}
\mL^{(0)}_{\mathrm{EFT}} = \frac{\pi\alpha_s}{2m_{\ti}^3}
(\mY_n \mB^{\mu}_{n\perp})^a (\mY_{\n} \mB^{\perp}_{\n\mu})^b
\Bigl[\frac{\delta_{ab}}{N_c} \tiv^{\dagger}\timv + \mY_v^{\dagger km}D^k_{ab} \tiv^{\dagger} T^m \timv\Bigr].
\ee

\section{Factorization of stop pair production near partonic threshold}

Near the partonic threshold for $gg \to \ti^{\dagger} \ti$, the scattering cross section for $pp \to \ti^{\dagger} \ti X$ can be factorized into hard, soft,
 collinear and Coulombic parts. The derivation of the factorization theorem
is similar to the one obtained in Ref.~\cite{Idilbi:2010rs}, where the production of
color-octet scalar pairs is studied. The factorization theorem for the final state with color representation $R_f$ is
\bea
\sigma_{R_f} (pp \to \ti\ti^{\dagger}X) &=&  \int d x_1 dx_2 d\eta \sum_{R_i} \frac{|C_{R_i,R_f}(M,\mu)|^2}{8m_{\ti}^6 (N_c^2-1)^2} \hat{s} \tilde{f}_{g/p} (x_1) \tilde{f}_{g/p} (x_2) \nonumber \\
\label{fact1}
&&\times S_{R_i,R_f} (\eta)~ \mathrm{Im}\, G_{R_f} (0,0,E+i\Gamma_{\ti}).
\eea
Here $E = \hat{s}^{1/2} - 2m_{\ti} -\eta = M -2m_{\tilde t}$, where $M$ is the invariant mass of the stop pair and $\eta$ is given by $p_{X_S}^0$
in the CM frame for the incoming partons, i.e., $\eta$ is the total energy carried by soft particles in the CM frame.
We note that $M$ is approximately equal to the stoponium mass since the difference is negligible when  $M$ is considered as the highest available scale and all other scales are small compared to it. Thus, in our analysis, we do not distinguish between the stoponium mass and the invariant mass of the stop pair.
$G_{R_f}(0,0, E) $ is the Green's function for the final $R_f$ state, which describes Coulombic interactions between the
 heavy stop pair.
In our work, we use the Green's function computed to NLO in $\alpha_s$. The explicit formulas are given in Eqs.~(\ref{GFLO}) and (\ref{GFNLO}) of Appendix C.
$\Gamma_{\tilde t}$ is stop decay rate.

Before we continue the derivation of the factorization theorem we pause for a while and discuss the possibility of bound state formation.
The SUSY scenario we are focusing on is where stop is the NLSP
and the stop mass is less than the sum of the top mass and neutralino mass as well as the sum of the bottom and chargino mass so the tree-level two-body decays of stop are forbidden. In order for this scenario to be realized we must have $m_{\chi^0} <  m_{\tilde t} < m_b + m_{\chi^+}$
(where $m_{\chi^+}$ and $m_{\chi^0}$ are the masses of the lightest chargino and neutralino, respectively). Therefore, the neutralino and chargino cannot be degenerate. This type of SUSY spectrum can be obtained by relaxing the ``natural SUSY'' requirement $M_1,M_2 \gg |\mu|$.
In this scenario, the main stop decay channels are loop-induced charm
quark and neutralino decay and three-body and four-body cascade decays. For these decay
channels, $\Gamma_{\tilde t}$ is a few keV or smaller~\cite{Hikasa:1987db,Porod:1996at}. Comparing this width with the binding energy of stoponium which is $1$-$3 \GeV$, we see that the stop pair will live long enough to form the stoponium before they decay. Since $\Gamma_{\tilde t} \ll m_{\tilde t}$  finite-width effects are negligible~\cite{Falgari:2012sq}.

$\tilde f_{g/p}$ is the collinear function for the gluon field, which is matched  onto the standard parton distribution function (PDF).
The soft function $S_{R_i,R_f}(\eta)$ is defined as
\bea
\label{softf}
S_{R_i,R_f} (\eta) &=& \sqrt{\mathrm{dim} R_i}~ E_{ab\alpha\beta}^{(R_i,R_f)*}E_{cd\gamma\delta}^{(R_i,R_f)}E_{pqrs}^{(R_f,R_f)*} \\
&&\times \langle 0 | \mY_n^{\dagger ea} \mY_{\n}^{\dagger fb} Y_{v,\alpha p}^{\dagger} Y_{v,q\beta} \delta(\eta+ i\partial_0) \mY_n^{ce} \mY_{\n}^{df} Y_{v,r\gamma}  Y_{v,\delta s}^{\dagger} | 0 \rangle. \nonumber
\eea
At tree level we have: $S_{R_i,R_f}^{(0)}(\eta) = \delta (\eta)$.

If we use the variable $z= M^2/\hat{s} = \tau/(x_1 x_2) \sim 1$ where $\tau = M^2/s$ and $s$ is the CM energy of the incoming two protons, the soft
 momentum $\eta$ can be written as
\be
\label{softm}
\eta = \hat{s}^{1/2} - M = \hat{s}^{1/2} (1-z^{1/2}) \sim \frac{M}{2}(1-z).
\ee
The differential scattering cross section as a function of the invariant mass $M$ is
\bea \label{dfact1}
\frac{d\sigma_{R_f}}{dM} &=& \sum_{R_i} H_{R_i,R_f} (M,\mu) \frac{M}{(2m_{\ti})^6} \mathrm{Im}\, G_{R_f} (0,0,M-2m_{\ti}+i\Gamma_{\ti}) \\
&&\times~\tau \int^1_{\tau} \frac{dz}{z} \bar{S} _{R_i,R_f} (1-z,\mu) \tilde F_{gg} \Big(\frac{\tau}{z},\mu\Big), \nonumber
\eea
where the hard function $H_{R_i,R_f}(M,\mu)$ is given by
\be \label{hardf}
H_{R_i,R_f} (M,\mu) = 16 \frac{|C_{R_i,R_f}(M,\mu)|^2}{(N_c^2-1)^2}.
\ee
The function $\tilde F_{ij}(x,\mu)$ is the convolution  of two collinear functions from $i,j$ initial partons:
\be
\label{convf}
\tilde F_{ij} (x,\mu) = \int^1_x \frac{dy}{y} \tilde f_{i/p} (y,\mu) \tilde f_{j/p} (x/y,\mu),
\ee
and the dimensionless soft function in Eq.~(\ref{dfact1})
$\bar{S}_{R_i,R_f}(1-z) = (M/2) S_{R_i,R_f} (\eta)$ is normalized so that $\bar{S}_{R_i,R_f}^{(0)}(1-z) = \delta(1-z)$.

The soft function $\bar{S}_{R_i,R_f}(1-z)$ as well as $S_{R_i,R_f} (\eta)$ are infrared (IR) divergent. Hence Eq.~(\ref{dfact1}) cannot describe
the same low energy physics as full QCD if the collinear function $\tilde{f}_{g/p}$ is a genuine PDF. Recently it was pointed out that one has to
subtract the contribution of the mode $p_s \sim Q(1-z)$ from the collinear function in order to avoid double-counting problems between
 the collinear and soft parts~\cite{Chay:2012jr,Chay:2013zya}\footnote{This subtraction is done partonically and order by order in perturbation theory.}. Then the collinear function can be matched onto the PDF.
The gluonic collinear function can be written as the convolution of the collinear kernel and the gluon PDF~\cite{Chay:2013zya,Fleming:2006cd}
\be\label{colm}
\tilde f_{g/p} (x,\mu) = \int^1_x \frac{dz}{z} K_{gg} (z,\mu) f_{g/p} \Big(\frac{x}{z},\mu\Big),
\ee
where $K_{gg}(z,\mu)$ is the collinear kernel and $f_{g/p}$ is the gluon PDF. When combining the soft function with the two collinear kernels we obtain an
 IR finite kernel
\be\label{Wgg}
W_{R_i,R_f} (1-w,\mu) = \int_w^1
\frac{dz}{z}  \bar{S}_{R_i,R_f} (1-z,\mu) \int_{w/z}^1 dt K_{gg} (t,\mu)
K_{gg} \Bigl(\frac{w}{zt},\mu\Bigr).
\ee
Putting Eqs.~(\ref{colm}) and (\ref{Wgg}) into Eq.~(\ref{dfact1}), we rewrite the differential scattering cross section as
\bea \label{dfact2}
\frac{d\sigma_{R_f}}{dM} &=& \sum_{R_i} H_{R_i,R_f} (M,\mu) \frac{M}{(2m_{\ti})^6} \mathrm{Im}\, G_{R_f} (0,0,M-2m_{\ti}+i\Gamma_{\ti}) \\
&&\times~\tau \int^1_{\tau} \frac{dz}{z} W_{R_i,R_f} (1-z,\mu) F_{gg}\Big(\frac{\tau}{z},\mu\Big), \nonumber
\eea
where $F_{ij}(x,\mu)$ is the parton luminosity function with initial partons $i,j$,
\be
\label{convpdf}
F_{ij}(x,\mu) = \int^1_{x} \frac{dy}{y} f_{i/p} (y,\mu)f_{j/p} \Big(\frac{x}{y},\mu\Big).
\ee
In general, the factorization scale $\mu_F$ in Eq.~(\ref{dfact2}) should be considered to be smaller than the intermediate scale $\mu_S \sim M(1-z)$
since we have successively
 integrated out the hard ($\sim M$) and the soft ($\sim M(1-z)$) modes in order to obtain Eq.~(\ref{dfact2}).

Changing the variable $y$ in Eq.~(\ref{convpdf}) to the rapidity of the stop pair, $Y$, we have the following doubly differential scattering cross section
\bea \label{dfact3}
\frac{d\sigma_{R_f}}{dMdY} (pp \to \ti\ti^{\dagger}X) &=& \sum_{R_i} H_{R_i,R_f} (M,\mu) \frac{M}{(2m_{\ti})^6} \mathrm{Im}\, G_{R_f} (0,0,M-2m_{\ti}+i\Gamma_{\ti}) \\
&&\times~\tau \int^1_{\tau} \frac{dz}{z} W_{R_i,R_f} (z,\mu) f_{g/p} \Big(\frac{Me^Y}{\sqrt{s}},\mu\Big)
f_{g/p} \Big(\frac{Me^{-Y}}{\sqrt{s}},\mu \Big), \nonumber
\eea
where we ignored soft momentum contributions to the rapidity since they are subleading.

\section{Scattering Cross section for $pp \to {\tilde \sigma}$ and resummation}

Near threshold, the produced stop pair moves slowly enough to form a stoponium bound state, $\tilde \sigma$.
 The produced bound state can decay to electroweak gauge bosons such as $\gamma\gamma$, $\gamma Z$, $ZZ$, and $W^+W^-$.
In the case of bound states of color-octet scalars, the signal for pairs of electroweak bosons can exceed the SM background at the LHC~\cite{Kim:2008bx,Idilbi:2010rs}. For the stop pair production,
the signal might be weak compared to the color-octet scalar because of the relatively small Casimir factor $C_F$. However, with
sufficient integrated luminosity, we will see that the signal for stoponium can be visible above SM backgrounds, especially at the 14 TeV energy of future LHC runs. In this section we study the scattering cross section
 for stoponium production followed by its electroweak decays, $pp \to \tilde \sigma \to AB$. The stops and the stoponium are very narrow in the scenario we are considering, so we expect the cross section to be enhanced
in a narrow region around $M \sim 2 m_{\ti}$. Since the decay width of the stoponium is a few tens of MeV, we use Eq. (\ref{dfact3}) multiplied by the  branching ratio
for $\tilde \sigma \to AB$  in order to obtain the cross sections $pp \to \tilde \sigma \to AB$  in our analysis.
Since we consider electroweak decays of the stoponium, we only consider color singlet production and provide the relevant radiative corrections and resummation of large logarithms.

Combining Eqs.~(\ref{wilt}) and (\ref{hardf}) we obtain the LO contribution to the hard function in Eq.~(\ref{dfact3}) (for $R_i =R_f ={\bf 1}$),
\be\label{hardlo}
H_{\bf 1,1}^{(0)} (M,\mu) = \frac{16\pi^2 \alpha^2_s (\mu)}{N_c(N_c^2-1)}.
\ee
 Up to NLO the hard function $H_{\bf 1,1}(M,\mu)$ can be extracted from Ref.~\cite{Younkin:2009zn}
\be\label{hardnlo}
H_{\bf 1,1}(M,\mu) = H_{\bf 1,1}^{(0)} (M,\mu) \bigg[ 1+ \frac{\alpha_s (\mu)}{\pi} \Bigl\{C_A\Bigl(1+\frac{\pi^2}{3} - \frac{1}{2} \ln^2\frac{\mu^2}{M^2} \Bigr) - C_F \Bigl(3+ \frac{\pi^2}{4}\Bigr)\Bigr\} +\ldots \bigg],
\ee
and $C_A = N_c$ and $C_F = \frac{N_c^2-1}{2N_c}$.  The anomalous dimension of the term in the square parentheses is the same as the anomalous dimension for the hard
 scattering coefficient for Higgs boson production (via gluon-gluon fusion) since the effective theory calculations are identical at hard matching scale.
 Therefore, we can obtain the two-loop anomalous dimension of the hard function from the known result for Higgs boson production.
 The anomalous dimension of the hard function is given by
\be \label{anomalH}
\hat\gamma^H (\mu) = \frac{1}{H_{\bf 1,1}} \frac{d}{d\ln\mu} H_{\bf 1,1}
= 2 \Gamma_C^A(\alpha_s) \ln \frac{M^2}{\mu^2} + 2 \gamma^S + \frac{2 \beta(\alpha_s)}{\alpha_s}\,.
\ee
The cusp anomalous dimension $\Gamma_C^A(\alpha_s)$ in the adjoint representation and the anomalous dimension of the hard function for Higgs production $\gamma^S$ are perturbatively calculable. We parametrize their expansion in $\alpha_s$ as
\bea
\Gamma_C^A &=& \sum_{k=0} \Gamma_{C,k}^A\Big(\frac{\alpha_s}{4\pi} \Big)^{k+1}\,, \nn \\
\gamma^S &=& \sum_{k=0} \gamma_k^S\Big(\frac{\alpha_s}{4\pi} \Big)^{k+1}\,.
\eea
The coefficients of the cusp anomalous dimension up to three-loop order and the anomalous dimension of the hard factor up to two-loop order
are given in  Appendix B. The function $\beta(\alpha_s)$ is defined by
\be
\beta(\alpha_s) = \frac{d\alpha_s}{d\ln\mu}.
\ee
The expansion of $\beta(\alpha_s)$ begins at ${\cal O} (\alpha_s^2)$. We will need the  three-loop expression for $\beta(\alpha_s)$ in this work, and it is
 given in Appendix B.

The logarithms of the hard function in Eq.~(\ref{hardnlo}) are minimized at $\mu \sim M$. Hence we can identify the typical
 hard scale as $\mu_H \sim M$ for a stable perturbative expansion. However, if the factorization scale $\mu_F$ is taken to be much smaller than $\mu_H$,
 we must evolve the  hard function from the scale $\mu_H$ to the scale $\mu_F$. Using $\hat \gamma^H$ in Eq.~(\ref{anomalH}) we find
\bea \label{RGH}
H_{\bf 1,1}(M,\mu_F) &=& \bigg(\frac{\alpha_s(\mu_F)}{\alpha_s(\mu_H)}\bigg)^2 {\rm exp}\Big[ - 4S_\Gamma(\mu_F,\mu_H) + 2a_{\gamma^S}(\mu_F,\mu_H) \Big]\,\nn \\
&& ~ \Big(\frac{\mu_F^2}{M^2}\Big)^{-2a_\Gamma(\mu_F,\mu_H)}H_{\bf 1,1}(M,\mu_H)\,.
\eea
The Sudakov exponent $S_{\Gamma}(\mu_1,\mu_2)$ and the exponent $a_{\gamma^A} (\mu_1,\mu_2)$ for an arbitrary anomalous dimension $\gamma^A$ are defined by
\bea
\label{SF}
S_{\Gamma}(\mu_1,\mu_2) &=& \int^{\alpha_s (\mu_1)}_{\alpha_s (\mu_2)} \frac{d\alpha}{\beta(\alpha)}
\Gamma_C^A (\alpha) \int^{\alpha}_{\alpha_s (\mu_1)}  \frac{d\alpha'}{\beta(\alpha')}, \\
a_{\gamma^A} (\mu_1,\mu_2)  &=& \int^{\alpha_s (\mu_1)}_{\alpha_s (\mu_2)} \frac{d\alpha}{\beta(\alpha)}
\gamma^A(\alpha)\,,
\eea
and similarly $a_\Gamma(\mu_1,\mu_2)$ is defined by replacing $\gamma^A(\alpha)$ with $\Gamma_C^A(\alpha)$ in the definition of $a_{\gamma^A}(\mu_1,\mu_2)$.
The solutions for the Sudakov exponent and $a_{\Gamma}(\mu_1,\mu_2)$ up to NNLL order are given in Appendix B.

The soft kernel at NLO was computed in Ref.~\cite{Chay:2013zya} and is given by
\bea \label{softk}
W_{\bf1,1} (z,\mu) &=& \delta(1-z) \Bigl[1+\frac{\alpha_s C_A}{\pi} \Bigl(\frac{1}{2} \ln^2 \frac{M^2}{\mu^2} -\frac{\pi^2}{4} \Bigr) \Bigr] \\
&&+\frac{\alpha_s C_A}{\pi} \Bigl[2\ln\frac{M^2}{\mu^2} \frac{1}{(1-z)_+} + 4 \Bigl(\frac{\ln(1-z)}{1-z}\Bigr)_+ \Bigr]\, , \nonumber
\eea
and obeys the following renormalization group (RG) equation,
\be
\label{RGW}
\frac{d}{d\ln \mu} W_{\bf 1,1} (x,\mu) = \int^1_x \frac{dz}{z} \hat \gamma^W (z,\mu) W_{\bf 1,1}\Big(\frac{x}{z},\mu\Big) \, ,
\ee
where the anomalous dimension $\hat \gamma_W$ is
\be\label{anomalWG}
\hat\gamma^W(z,\mu) = - \left(2\Gamma_C^A(\alpha_s)\ln\frac{M^2}{\mu^2} + 2\gamma^W \right)\delta(1-z)
- \frac{4\Gamma_C^A(\alpha_s)}{(1-z)_+}\, .
\ee
Here $\gamma^W = 0 + \mO(\alpha_s^2)$.
One can also show that $\gamma^W = \frac{\beta(\alpha_s)}{\alpha_s}+ 2\gamma^B + \gamma^S$, where $2\gamma^B$ is the coefficient of the $\delta(1-x)$ term in the Altarelli-Parisi splitting function, $P_{gg}(x)$,  by demanding that Eq.~(\ref{dfact3})
is scale independent.

Solving the RG equation in Eq.~(\ref{RGW}) by applying  the Laplace transform~\cite{Becher:2006nr,Becher:2006mr}, we evolve $W_{\bf 1,1}$ from the soft
scale $\mu_S$ to the factorization scale $\mu_F$ using the formula
\bea \label{evoW}
W_{\bf1,1} (z,\mu_F) &=& \bigg(\frac{\alpha_s(\mu_S)}{\alpha_s(\mu_F)}\bigg)^2 \exp\Bigl[4S_{\Gamma} (\mu_F,\mu_S)-4 a_{\gamma^B} (\mu_F,\mu_S)-2a_{\gamma^S} (\mu_F,\mu_S)\Bigr]
 \nonumber \\
&&\times \Bigl(\frac{\mu_F}{M}\Bigr)^{-\eta} \tilde{w}_{\bf 1,1} \Big[\ln\frac{\mu_S}{M} -\frac{\partial_{\eta}}{2}\Big] \frac{e^{-\gamma_E \eta}}{\Gamma(\eta)} (1-z)^{-1+\eta}\,,
\eea
where $\eta$ is defined as $\eta = -4 a_{\Gamma}(\mu_F,\mu_S)$ and is positive for $\mu_F<\mu_S$.
$\tilde{w}_{\bf 1,1}(L)$ is obtained by substituting $L = \ln (\mu s e^{\gamma_E}/M)$ in $\tilde{W}_{\bf 1,1} (s)$:
\be
\tilde{w}_{\bf 1,1}(L) = \tilde{W}_{\bf 1,1} \Big( \frac{M}{\mu} e^{L-\gamma_E} \Big)\,,
\ee
where $\tilde{W}_{\bf 1,1}(s)$ is the Laplace transform of $W_{\bf 1,1}(z)$ in momentum space.  $\tilde{W}_{\bf 1,1}(s)$ is defined by
\be\label{LapW}
\tilde{W}_{\bf 1,1} (s) = \int^{\infty}_0 dt e^{-st} \hat{W}_{\bf 1,1} (t) = \int^1_0 dz z^{-1+s} W_{\bf 1,1}(z),~~~t=-\ln z,
\ee
where $\hat{W}_{\bf 1,1} (t) = W_{\bf 1,1}(z)$. Taking the limit $s\to \infty~(t\to 0)$, we compute
$\tilde{W}_{\bf 1,1} (s)$ at NLO in $\alpha_s$ to be
\be \label{LapWnlo}
\tilde{W}_{\bf 1,1} (s)  = 1 + \frac{\alpha_s C_A}{2\pi} \Bigl[\ln^2\frac{\mu^2 s^2e^{2\gamma_E}}{M^2}+\frac{\pi^2}{6} \Bigr]\,,
\ee
which leads to
\be
\tilde{w}_{\bf 1,1}(L) = 1 + \frac{\alpha_s C_A}{4\pi} \bigg[ 8 L^2 + \frac{\pi^2}{3} \bigg]\,.
\ee
Putting all the pieces  together we obtain
\bea
\label{eq:resummation}
H_{\bf 1,1} (M,\mu_F) W_{\bf 1,1}(z,\mu_F) &=& H_{\bf 1,1} (M,\mu_H) \bigg(\frac{\alpha_s(\mu_S)}{\alpha_s(\mu_H)}\bigg)^2
\Bigl(\frac{M}{\mu_H}\Bigr)^{-4a_\Gamma(\mu_H,\mu_S)} \nonumber \\
&& \times \exp\Bigl[4S_{\Gamma} (\mu_H,\mu_S)-4 a_{\gamma^B} (\mu_F,\mu_S)-2a_{\gamma^S} (\mu_H,\mu_S)\Bigr]
 \nonumber \\
&&\times  \tilde{w}_{\bf 1,1} \Big[\ln\frac{\mu_S}{M} -\frac{\partial_{\eta}}{2}\Big] \frac{e^{-\gamma_E \eta}}{\Gamma(\eta)} (1-z)^{-1+\eta}\,,
\eea
where we used the following relation,
\be
S_\Gamma(\mu_F,\mu_S) - S_\Gamma(\mu_F,\mu_H) = S_\Gamma(\mu_H,\mu_S) - a_\Gamma(\mu_H,\mu_S) \ln \frac{\mu_F}{\mu_H}\,.
\ee
The above resummation formula includes all order resummation of large
logarithms of $\ln\mu_H/\mu_S$. Treating $\alpha_s \ln (\mu_H/\mu_S)$ as ${\cal O}(1)$, we note that the expansion of
 $S_\Gamma(\mu_H,\mu_S)$ begins with $\alpha_s \ln^2 \mu_H/\mu_S \sim {\cal O}(1/\alpha_s)$
 at one-loop order while the expansion of $a_{\gamma^S} (\mu_F,\mu_S)$
 begins with $\alpha_s \ln \mu_F/\mu_S \sim {\cal O}(1)$ at one-loop order.
Therefore, in order to obtain a resummed cross section with the same accuracy as the
 NLO hard scattering contribution, we need  $S_\Gamma(\mu_H,\mu_S)$ to three-loop order and
  $a_{\gamma^{S(B)}} (\mu_F,\mu_S)$ to two-loop order. Doing this,  we achieve NNLL resummation accuracy.
All the ingredients that are needed for NNLL resummation are given in Appendix B.

\section{Phenomenology}
In this section we carry out a phenomenological analysis of stoponium production and decay focusing on the processes $pp\to\tilde{\sigma}\to\gamma\gamma$ and $pp\to\tilde{\sigma}\to ZZ$ relevant to  current accumulated as well as future LHC data. The $\gamma\gamma$ and $ZZ$ channels are golden modes  for searching for stoponium  simply because their invariant masses  can be cleanly reconstructed from the energy deposition in calorimeters and the momentum of tracks of their decay products in the collider detector. In addition, SM backgrounds for $\gamma\gamma$ and $ZZ$ channels are much smaller compared with the $gg$ channel. The $Z\gamma$ channel is not favored since the branching ratio for $\tilde \sigma \to Z\gamma$ is much smaller than the branching ratios for the $\gamma\gamma$ or $ZZ$  channels, while the SM background is larger than the $ZZ$ channel.

It is interesting to search for resonances in the $WW$ channel. This has been done at the LHC by reconstructing $WW$ from two merged jets using jet substructure techniques~\cite{Chatrchyan:2012ypy,CMS:2013fea}. However, this analysis is beyond the scope of this work and we leave it for future work.

We choose a typical MSSM parameter set which is denoted by {\bf P1}: $\theta_{\tilde t} = \pi/4$, $\tan\beta=10$, $m_A = 2\TeV$ and $\kappa =-2$. Here, $\theta_{\tilde t}$ is the mixing angle between left and right stops where its typical value is chosen by maximal mixing, $\tan\beta$ is the ratio of the vacuum expectation value (VEV) of the up-type Higgs to the VEV of the down-type Higgs, and $m_A$ is the mass of CP odd neutral Higgs. The mixing angle, $\alpha$, between the neutral Higgs is obtained through the well-known relation  $\tan2\alpha=\tan2\beta(m_A^2+m_Z^2)/(m_A^2-m_Z^2)$.  $\kappa$ originates from the triple scalar coupling $\lambda_{ \tilde{t} \tilde{t} h}$ and is defined as in Ref.~\cite{Barger:2011jt} by (see Appendix A for more details)
 \begin{equation}
\kappa \,m_W = (-\mu \sin\alpha + a_t \cos\alpha)\,.
\end{equation}
Here, $\mu$ is the Higgsino mass  and $a_t$ is the trilinear coupling of scalars in the soft breaking term. The light Higgs mass is fixed by recent measurements to be $m_h=126\GeV$~\cite{Aad:2012tfa,Chatrchyan:2012ufa}. We set the mass of the light stop as a free parameter within the range of $ 200\GeV < m_{\tilde t} < 400\GeV$, for reasons discussed in the Introduction. Throughout this work, we neglect the contributions of the heavier stop, gluino and heavy Higgs in the intermediate state for stoponium production and decays by assuming that they are much heavier than the light stop.  As for the SM parameters, we use $\alpha_s(M_Z) = 0.117$, $m_t = 173.5\GeV$. For numerical analysis, we employed the MSTW2008NNLO PDF set \cite{oai:arXiv.org:0901.0002}.
In order to see uncertainty from choosing different PDF sets, we simulated heavy Higgs production comparing the results by using the CTEQ5, CTEQ6, and CTEQ10 PDF sets \cite{Lai:1999wy} as well as the MSTW2008NNLO PDF set. We find that the differences are always less than 5\%.

\begin{figure}[h]
\centering
\includegraphics[width=3.8in]{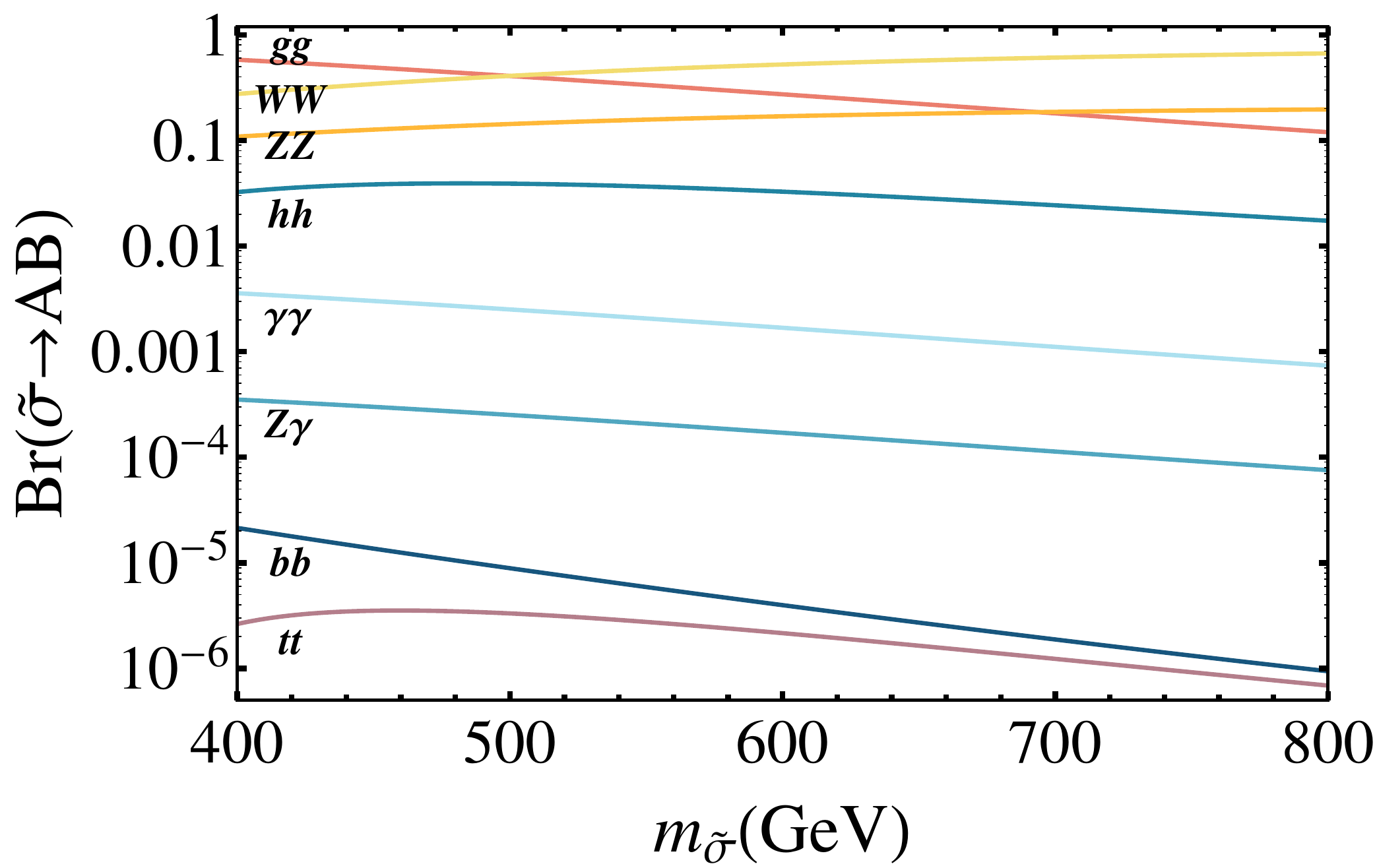}
\caption{
\baselineskip 3.0ex
Branching ratios of stoponium decays for each decay channel. MSSM parameters are chosen from set {\bf P1} (see the text). }
\label{fig:BR}
\end{figure}

With this choice of parameters, we plot  the branching ratios for two-body decays of stoponium as a function of the stoponium mass  in figure \ref{fig:BR}. The  stoponium decays are calculated at tree level  and formulas for stoponium decay rates are given in Appendix A. We have  confirmed that our results are analytically consistent with Refs.~\cite{Drees:1993uw,Martin:2008sv}, and numerically consistent with Ref.~\cite{Barger:2011jt}.  Here, we neglect the stoponium decay into neutralino pairs which is highly suppressed compared with the leading decay channel~\cite{Barger:2011jt,Drees:1993uw,Martin:2008sv}. As shown in the figure, the branching ratios for the  $WW$ and $ZZ$ channels increase  with increasing the stoponium mass  while other decay modes exhibit the opposite behavior.  To physically understand this property we note that the sum of the polarization vectors for a massive gauge boson is $\sum \epsilon_\mu \epsilon_\nu^* = - g_{\mu\nu} + k_\mu k_\nu / m_{Z,W}^2$ where  $k_\mu$ is the four-momentum of the massive gauge boson. The second term comes from  the longitudinal polarizations and becomes larger as the stoponium mass increases. Thus the branching ratios for $WW$ and $ZZ$ increase with the stoponium mass. We will discuss the dependence on MSSM parameters more in the last part of this section.

We next calculate the invariant mass distributions for stoponium decaying to both $\gamma\gamma$ and $ZZ$ channels using the factorization formula, Eq.~(\ref{dfact3}), multiplied by the appropriate branching fraction. We include the NLO hard function for stoponium production and the corresponding NNLL-order threshold resummation. Coulomb gluon resummation is taken into account using the NLO  Green's function. The LO Green's function is obtained by solving the Schr\"odinger equation for a $C_F \alpha_s/r$ potential, and hence is equivalent to resumming the leading-order Coulomb exchanges to all orders. We include this as well as NLO corrections.   The results for the cross section are shown with total scale uncertainty by adding in quadrature the errors associated with  variations of the factorization scale $\mu_F$,  hard scale $\mu_H$ and soft scale $\mu_S$. We discuss the scale choices and scale variations in more detail later in this section.

As for the stoponium production, we use RG-improved production cross section which is given by
\be
\hat\sigma^{\rm {RGI}}_{ij}(z) = \hat\sigma^{\rm {Res}}_{ij}(z) + \big( \hat\sigma^{\rm {Fixed}}_{ij}(z)|_{\mu_F}  - \hat\sigma^{\rm {Res}}_{ij}(z)|_{\mu_H = \mu_S = \mu_F}\big)
\ee
for initial-state patrons $ij$.  Here the terms in parentheses are expanded to NLO in $\alpha_s$ so we have the full NNLL resummed cross sections as well as the full NLO calculation
without double counting. The total cross section is
\be
\sigma = \sum_{ij} \int_\tau^1 \frac{dz}{z} \hat\sigma^{\rm {RGI}}_{ij}(z,\mu_H,\mu_S,\mu_F) \,\Phi_{ij}\Big(\frac{\tau}{z},\mu_F\Big)\,.
\ee
The differential cross section with respect to the invariant mass of stoponium can be obtained in a similar manner. The RG-improved cross section includes threshold resummation of the  terms that are singular as $z \rightarrow1$ as well as nonsingular contributions arising from real gluon emission into the final state. The partonic resummed cross section $\hat\sigma^{\rm {Res}}_{gg}(z)$ can be inferred from Eq.~(\ref{eq:resummation}):
\bea
 \hat\sigma^{\rm {Res}}_{gg}(z,\mu_H,\mu_S,\mu_F) &=& \frac{\pi}{\hat s}  \frac{|\psi(0)|^2}{M^3} H_{\bf 1,1} (M,\mu_H) \bigg(\frac{\alpha_s(\mu_S)}{\alpha_s(\mu_H)}\bigg)^2
\Bigl(\frac{M}{\mu_H}\Bigr)^{-4a_\Gamma(\mu_H,\mu_S)} \nonumber \\
&& \times \exp\Bigl[4S_{\Gamma} (\mu_H,\mu_S)-4 a_{\gamma^B} (\mu_F,\mu_S)-2a_{\gamma^S} (\mu_H,\mu_S)\Bigr]
 \nonumber \\
&&\times  \tilde{w}_{\bf 1,1} \Big[\ln\frac{\mu_S}{M} -\frac{\partial_{\eta}}{2}\Big] \frac{e^{-\gamma_E \eta}}{\Gamma(\eta)} (1-z)^{-1+\eta}\,.
\eea
Here, $\psi(0)$ is the stoponium bound state wave function at the origin, defined in the same way as in Ref.~\cite{Younkin:2009zn}.

 The procedure to get the invariant mass distribution of RG-improved cross section follows similar steps of previous section.
The NLO fixed-order calculation is separated into the part which is singular  as $z\to 1 $ and the other part which is regular up to $\ln(1-z)$ as $z\to 1$, namely, $\hat\sigma^{\rm {Fixed}}_{ij}(z) = \hat\sigma^{\rm {Sing}}_{ij}(z) + \hat\sigma^{\rm {Reg}}_{ij}(z)$.
The full SUSY-QCD correction to stop pair production at NLO was calculated in Ref.~\cite{Beenakker:1997ut}. In this work, we use the results of NLO QCD correction to stoponium production given in Ref.~\cite{Younkin:2009zn} while assuming that the gluino is much heavier than the light stop. The fixed NLO results are
\begin{eqnarray}
\hat\sigma^{\rm {Sing}}_{gg}(z) &=& \hat\sigma_0 \Bigg[ \delta(1-z)\bigg( 1 + \frac{\alpha_s}{\pi}(C_A - 3 C_F)\Big(1+\frac{\pi^2}{12}\Big) \bigg) + \frac{\alpha_s}{\pi}\bigg(2C_A \frac{1}{[1-z]_+} \ln \frac{ M^2}{\mu^2}
\nn \\ ~~~~~~~~~~~~~~~~~ &&
 + 4C_A \bigg[\frac{\ln(1-z)}{1-z}\bigg]_+ \bigg)\Bigg]\,,
\eea
\bea
\hat\sigma^{\rm {Reg}}_{gg}(z) &=& \hat\sigma_0 \frac{\alpha_s}{\pi} C_A \Bigg[   \frac{11z^5+11z^4+13z^3+19z^2+6z-12}{6z(1+z)^2 } -\frac{3}{1-z} \nn
 \\ ~~~~~~~~~~~~~~~~~ &&
+\frac{2(z^3-2z^2-3z-2)(z^3-z+2)z \ln z}{(1+z)^3(1-z)^2}
 \\ ~~~~~~~~~~~~~~~~~ &&
+2\Big(\frac{1}{z} + z(1-z)-2\Big) \ln\frac{M^2}{\mu^2} (1-z)^2  \Bigg]\,,\nn  \\
\hat\sigma^{\rm {Reg}}_{gq}(z) &=& \hat\sigma_0 \frac{\alpha_s}{\pi} \frac{C_F}{2}\Bigg[ 2+z-\frac{2}{z}-z\ln z + \frac{1+(1-z)^2}{z} \ln \frac{M^2}{\mu^2}(1-z)^2 \Bigg]\,, \\
\hat\sigma^{\rm {Reg}}_{q \bar q}(z) &=& \sigma_0 \frac{\alpha_s}{\pi} C_F^2 \frac{2}{3} z(1-z),
\end{eqnarray}
where the LO  cross section is given by
\begin{eqnarray}
\hat\sigma_0 = \frac{16\pi^3\alpha_s^2}{N_c (N_c^2-1) \hat s} \frac{|\psi(0)|^2}{M^3}\,.
\end{eqnarray}
For  $gq$ and $q\bar q$ at the initial state, there are no singular contributions at threshold.
One can also show that  $\hat\sigma^{\rm {Sing}}_{gg}(z)$ is reproduced by setting $\mu_H = \mu_S = \mu_F$ in the resummed cross section $\hat\sigma^{\rm Res}_{gg}(z,\mu_H,\mu_S,\mu_F)$ and expanding to $O(\alpha_s)$.

The decay rate of stoponium is a few tens of $\MeV$, therefore, a very narrow and sharp resonance signal is expected. However, in the experiments the resonant signals will be accumulated in a finite bin size that depends on  the resolution of the detectors. The ATLAS Collaboration reports that the expected photon energy resolution is \cite{ATLAS:2008}
\begin{equation}
\frac{\Delta E_\gamma}{E_\gamma} = \sqrt{\bigg(\frac{0.1}{E_\gamma/{\rm GeV}}\bigg)^2+0.007^2}\,
\end{equation}
for a detected photon energy $E_\gamma$. By roughly taking the photon energy $E_\gamma \approx m_{\tilde \sigma}/2 \leq 400 \GeV$ we obtain $\Delta E_\gamma \lesssim 2.8 \GeV$. We simply take $\Delta E = 2 \GeV$ as the bin size for the invariant mass distribution.
We define the resonant cross section of stoponium $\sigma_{\rm res}$ as an integral over the differential cross section within $\Delta E$ near the ground state resonant peak of stoponium:
\begin{eqnarray}
\sigma_{\rm res}^{AB} =  \int_{M_{\rm peak}-\frac{\Delta E}{2}}^{M_{\rm peak}+\frac{\Delta E}{2}} \frac{d\sigma(pp\to \tilde{\sigma} \to AB)}{dM} dM,
\end{eqnarray}
where $M_{\rm peak}$ denotes the invariant mass value where the first resonant peak arises.

The SM backgrounds are generated by the MCFM package \cite{Campbell:2011bn} for both $pp \to \gamma\gamma$ and $pp \to ZZ$ processes with NLO QCD correction.
The NLO correction to the $pp \to \gamma\gamma$ process includes the  one-loop $gg$ initial-state contribution. We use the following kinematical cuts:
\begin{eqnarray}
|\eta_{\gamma_{_{1,2}}}| < 2.4, ~~~~ p^T_{\gamma_{_{1,2}}} > 10 \GeV\,.
\end{eqnarray}
We note that the $p^T_\gamma$ cut has no impact for the large invariant mass region that we are focusing on when we apply the rapidity cut given above. We do not include secondary photons which come from the fragmentation of decaying partons. For the $ZZ$ channel, we computed the $ZZ$ invariant mass distribution for signal and SM background.  We did not multiply by branching ratios for the $Z$'s to decay to final states  with four leptons, two leptons and two jets, or four jets, which are actually observed in experiments.   We checked that the generated background is consistent with current experimental results in the low invariant mass region with the same kinematical cuts \cite{ATLAS:2012gamgam,ATLAS:2012ZZ}.

\begin{figure}[t]
\centering
\includegraphics[width=6.5in]{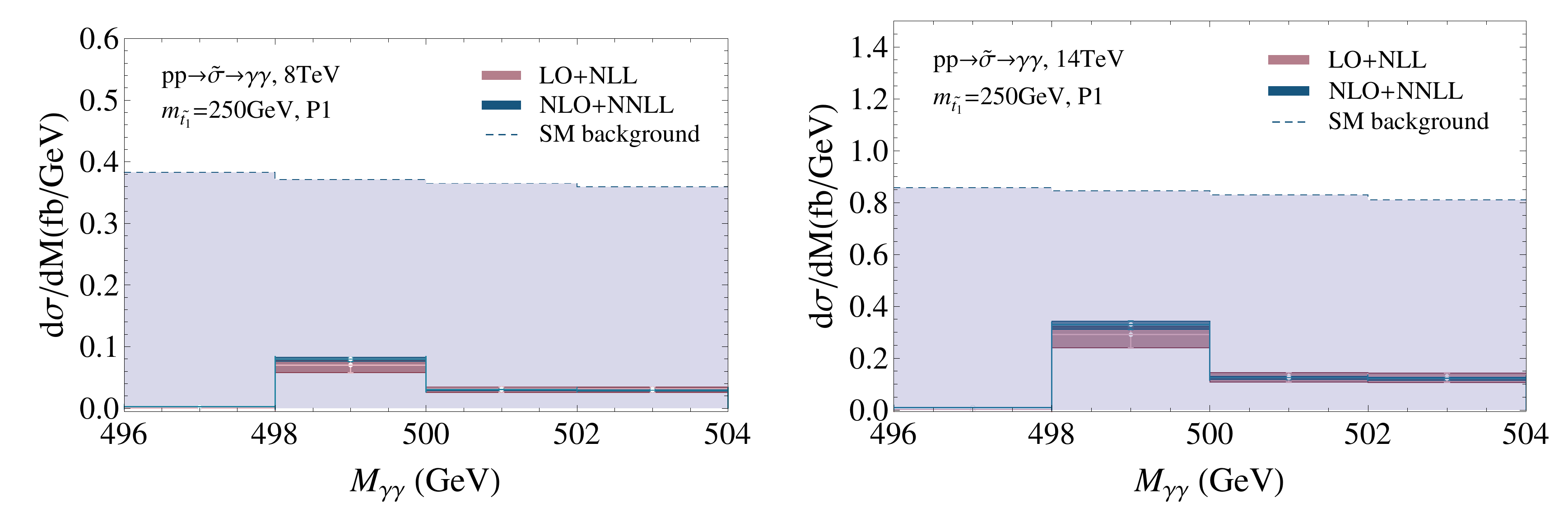}
\caption{
\baselineskip 3.0ex
$\gamma\gamma$ invariant mass distribution with a $2\GeV$ bin for both $pp\to\tilde{\sigma}\to\gamma\gamma$
signal and the SM background. Error bars represent total scale uncertainty. The MSSM parameter set is {\bf P1}. }
\label{fig:InvMgamgam}
\end{figure}
\begin{figure}[t]
\centering
\includegraphics[width=6.5in]{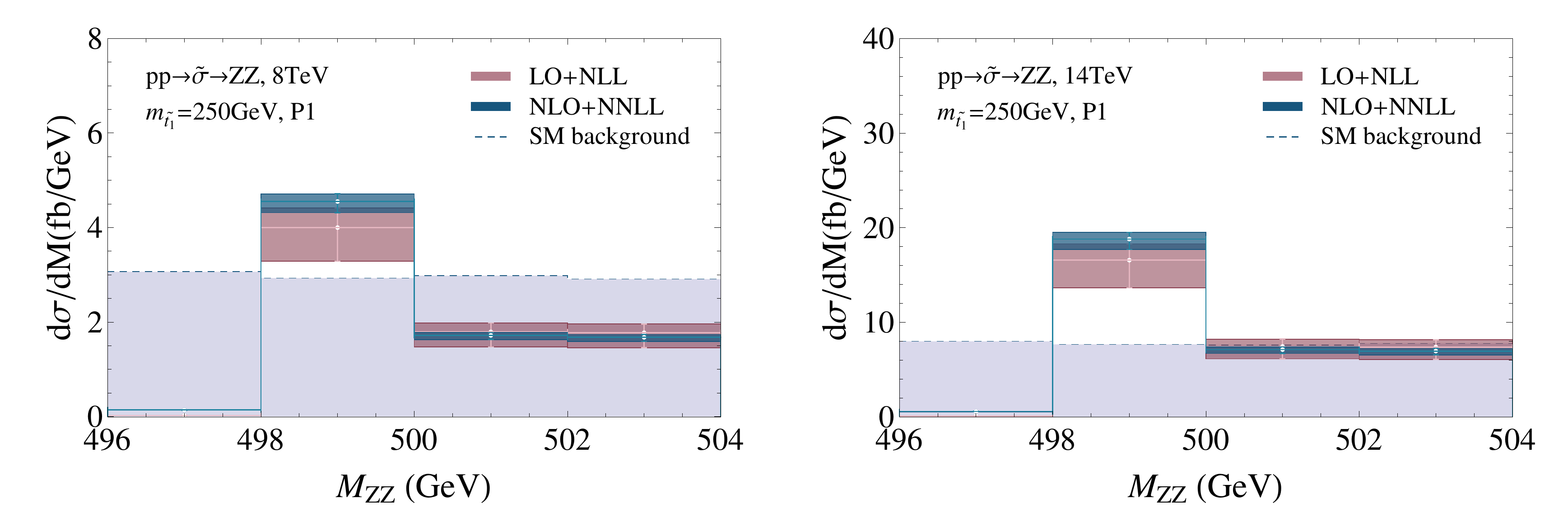}
\caption{
\baselineskip 3.0ex
$ZZ$ invariant mass distribution with a $2\GeV$ bin for both $pp\to\tilde{\sigma}\to ZZ$
signal and the SM background. Error bars represent total scale uncertainty.  The MSSM parameter set is {\bf P1}. }
\label{fig:InvMZZ}
\end{figure}
By setting the stop mass to $250\GeV$, we show a $2\GeV$-binned differential cross section of $pp\to\tilde{\sigma}\to\gamma\gamma$ as well as the SM background for both 8 and $14\TeV$ LHC runs. This is given in figure \ref{fig:InvMgamgam}. Each plot of the stoponium signal displays the total scale uncertainty in the error bars. It should be emphasized that the result shows good convergence of the perturbative expansion since the scale uncertainty is significantly reduced at NLO+NNLL. We note that the signal yield is much enhanced at $14\TeV$ as compared to $8\TeV$. The reason is that the $gg$ production channel is dominant for the signal while $q\bar q$  is dominant for the SM background, and the luminosity for initial-state $gg$ is much bigger than the luminosity of initial-state $q\bar q$ at higher center-of-mass energy. Therefore, with this parameter set, we can expect to see the stoponium signal at the early stages of the $14\TeV$ LHC run if it exists. In the last part of this section, we will give estimates  for the required luminosity for a $5\sigma$ discovery of the stoponium in future LHC runs.

The invariant mass distribution for $pp \to \tilde{\sigma} \to ZZ$ near the threshold region is shown in figure \ref{fig:InvMZZ}. The resonant signals are dominant over the SM background for both 8 and $14\TeV$. This is due to a much larger branching ratio for $\tilde{\sigma}\to ZZ$ than $\tilde{\sigma}\to \gamma\gamma$ in the MSSM parameter set {\bf P1}.
We note that if we take into account the $ZZ \to 4l\, (l = e, \mu)$ channel, both signal and background events will be reduced by factor of $0.0045$.
Nonetheless, in this parameter set, searching for the resonant signal in the $ZZ$ invariant mass distribution will serve as a  promising strategy for searching for stops.

\begin{figure}[t]
\centering
\includegraphics[width=6.5in]{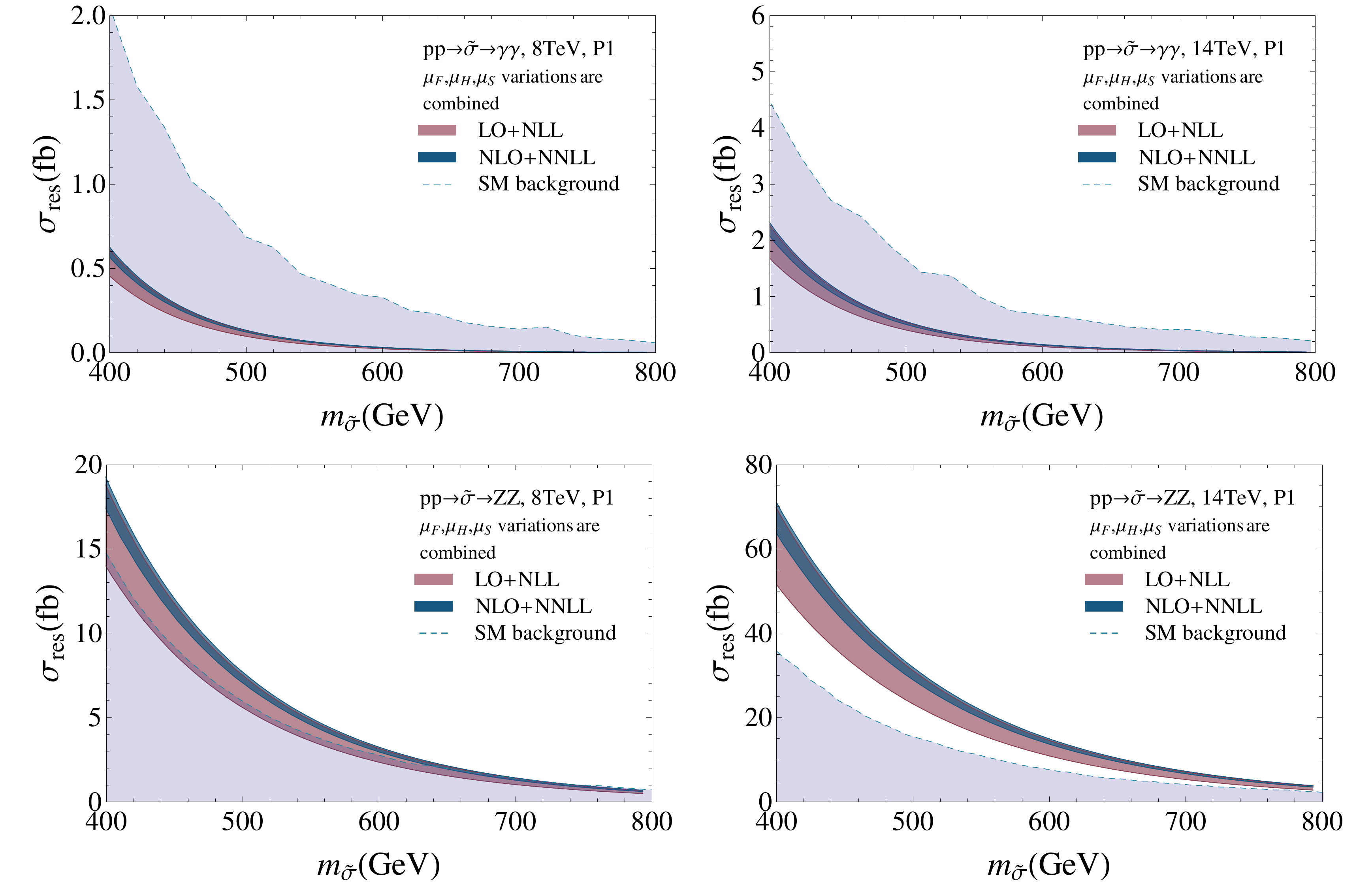}
\caption{
\baselineskip 3.0ex
Resonant cross section plot with respect to stoponium mass. Error bars represent total scale uncertainty. The MSSM parameter set is {\bf P1}. }
\label{fig:MstopDist}
\end{figure}
To address the issue of the stop search dependence on its mass, we vary, in our analysis, the stop mass parameter. Figure \ref{fig:MstopDist} shows the plots of $\sigma_{\rm res}^{\gamma\gamma}$ and $\sigma_{\rm res}^{ZZ}$ with respect to the stoponium mass for both 8 and $14\TeV$. The SM backgrounds are displayed in each plot for comparison. We again notice that the scale dependence is much reduced at NLO+NNLL for all stoponium masses. In all  cases, the $K$-factor is found to be  $1.09$ regardless of the stoponium mass. In the small mass region, the signal-to-background ratio is much enhanced. This is easily understood since the stoponium production rate is proportional to $1/m_{\tilde t}^3$ at LO. For $8\TeV$ in the $\gamma\gamma$ channel, the background is dominant for the entire stoponium mass range. In this case in order to find clear resonant signal we need large amount of data. Therefore, it is extremely difficult to find the $\gamma\gamma$ signal at $8\TeV$ with the currently accumulated data at the LHC since the cross section is small and there is a  poor signal to background ratio. However, the plot shows that the signal prevails over the background for most of the stoponium  mass range for both 8 and $14\TeV$ in the  $ZZ$ channel. Therefore, we anticipate that it may be possible to observe the  stoponium signal for stoponium masses below $800\GeV$ in the $14\TeV$ LHC run.  It should be noted that this result is obtained for the MSSM parameter set {\bf P1}, and there is significant MSSM parameter dependence.
 Before we go further into parameter dependence  we discuss  scale variation in our calculated cross sections.

\begin{figure}[t]
\centering
\includegraphics[width=6.5in]{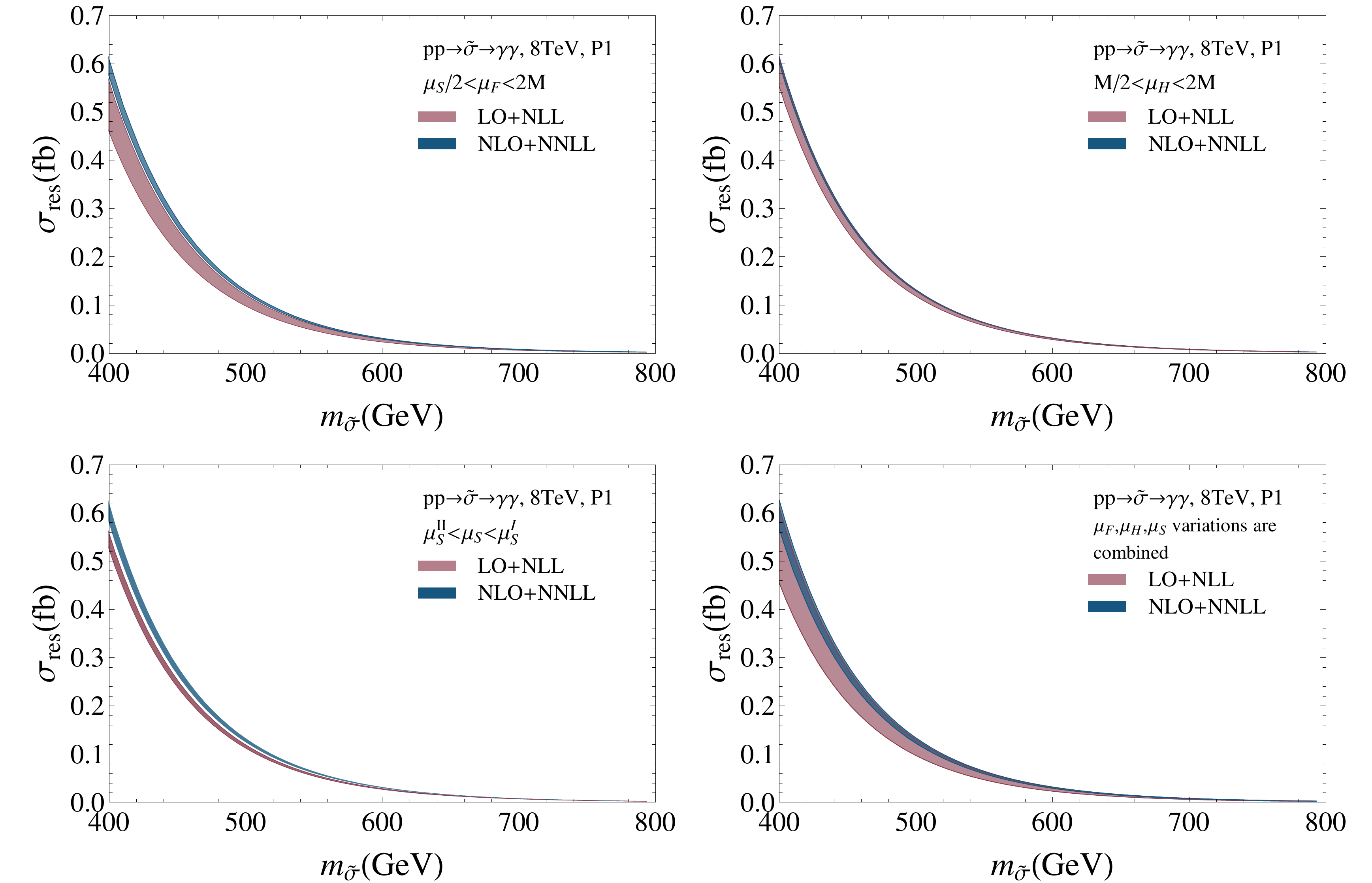}
\caption{
\baselineskip 3.0ex
Scale variations of resonant cross section with respect to stoponium mass. The MSSM parameter set is {\bf P1}. $\mu_S^I$ and $\mu_S^{II}$ are defined in the text. }
\label{fig:Scale}
\end{figure}

We choose the default value for the hard scale $\mu_H$ to be $\mu_H = M$, which suppresses large logarithms that can arise from the scale difference between $\mu_H$ and $M$.  We choose the default value of the factorization scale, $\mu_F$, to be the same as the default hard scale, $\mu_F = \mu_H = M$, which also suppresses large logarithms coming from large scale difference between $\mu_F$ and $\mu_H$. Even though one expects $\mu_F < \mu_S <\mu_H$ from the effective field theory point of view, in principle the cross section is independent of the scale chosen for $\mu_F$. In order to cover the region $\mu_F < \mu_S$ in the scale variation of $\mu_F$, we set the minimum variation of $\mu_F$ to be $\mu_S/2$. As we see below, the dependence on $\mu_F$   is very small in the NNLL resummed resonant cross section. We vary $\mu_F$ and $\mu_H$ in the following ranges
\begin{eqnarray}
  \mu_S/2 < \mu_F <  2M,~~~~ M/2 < \mu_H < 2 M\,,
\end{eqnarray}
where $\mu_S$ is default value of soft scale. The scale choice for  the soft scale $\mu_S$ is nontrivial. For this issue we follow Refs. \cite{Becher:2007ty,Ahrens:2008nc}. We define $\mu_S^I$ and $\mu_S^{II}$ as follows. $\mu_S^I$ is the soft scale when the soft one-loop correction decreases by 15\% starting from a high scale, $\mu_S^{II}$ is the soft scale when the soft one-loop correction has a minimum value. We average these two estimates of $\mu_s$ to obtain
 the default value of $\mu_S$ and vary $\mu_S$ as follows:
\begin{eqnarray}
\mu_S({\rm default}) = (\mu_S^I+ \mu_S^{II})/2, ~~~ \mu_S^{II} < \mu_S < \mu_S^{I}\,.
\end{eqnarray}
We plot the resonant cross section as a function of the stoponium mass and show the individual and combined scale variations
in figure \ref{fig:Scale}. At LO+NLL, the major bulk of uncertainty comes from factorization scale and hard scale uncertainties. However  both uncertainties are dramatically reduced at NLO+NNLL. It is remarkable that the factorization scale dependence is so small in the NNLL resummed cross section even though we vary $\mu_F$ in such a broad range. The soft scale uncertainty is also quite small at LO+NLL due to  large logarithmic resummation of $\ln(\mu_S/\mu_F)$. As mentioned before, the total scale uncertainty is greatly reduced at NLO+NNLL.

\begin{figure}[t]
\centering
\includegraphics[width=6.5in]{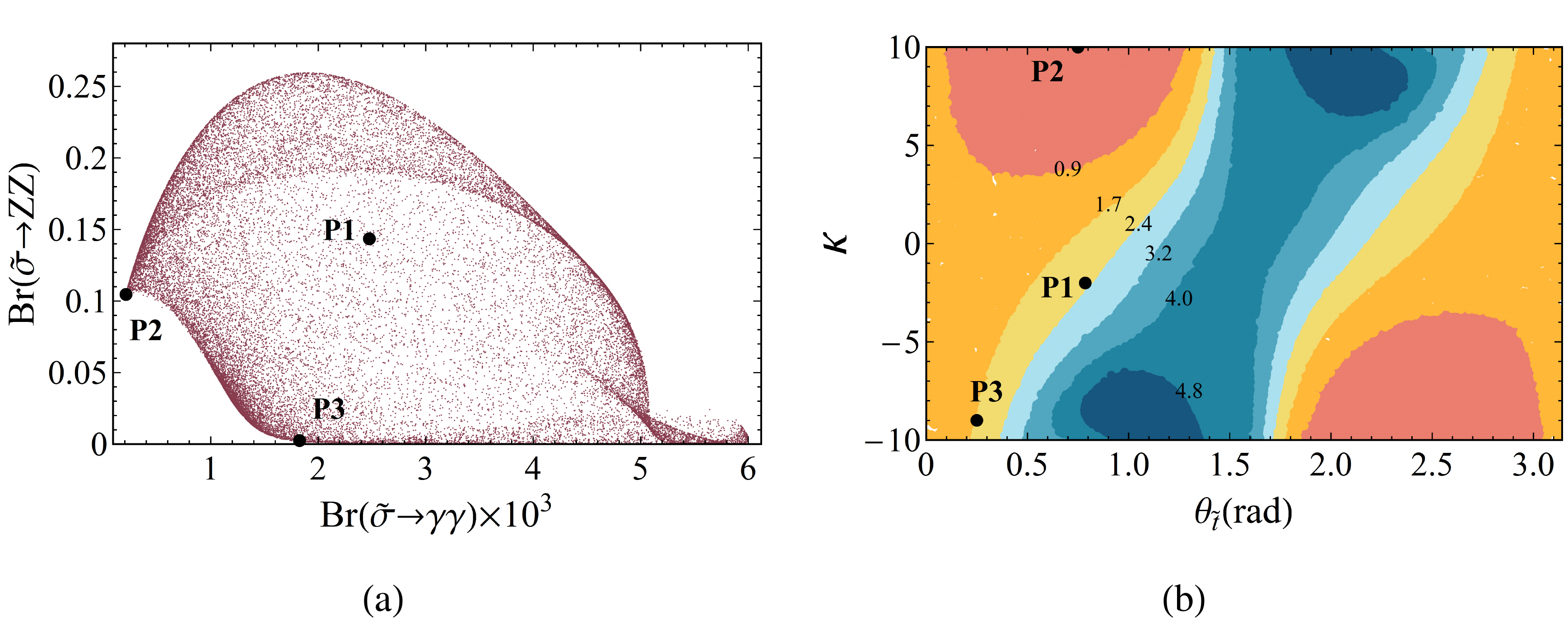}
\caption{
\baselineskip 3.0ex
(a) Scatter plot for ${\rm Br}(\tilde{\sigma}\to\gamma\gamma)$ and ${\rm Br}(\tilde{\sigma}\to ZZ)$ with respect to four relevant MSSM parameters. Here, $m_{\tilde t} = 250\GeV$. We choose three different benchmark points (see the text) : {\bf P1} (typical MSSM parameters), {\bf P2} (${\rm Br}(\tilde{\sigma}\to\gamma\gamma)$ is at its minimum), {\bf P3} (${\rm Br}(\tilde{\sigma}\to ZZ)$ is close to its minimum, with the minimum possible ${\rm Br}(\tilde{\sigma}\to\gamma\gamma)$ subject to this constraint).
 (b) Contour plot for ${\rm Br}(\tilde{\sigma}\to \gamma\gamma)\times 10^3$ in the $\theta_{\tilde{t}}-\kappa$ parameter space.
Three benchmark points {\bf P1}, {\bf P2} and {\bf P3} are shown in the figure.
}
\label{fig:MSSMparam}
\end{figure}
Now we study the dependence on the MSSM parameters for the resonant cross section. Since the resonant cross section is proportional to the branching ratio for the decay channel, it suffices to examine the MSSM parameter dependence of the stoponium branching ratios. We show the scatter plot for branching ratios for ${\tilde\sigma} \to \gamma\gamma$ and ${\tilde\sigma} \to ZZ$
in figure \ref{fig:MSSMparam}(a) with $m_{\tilde t} = 250\GeV$ by randomly generating the four relevant MSSM parameters within the ranges:
\begin{eqnarray}
&& ~~~~~ 0 < \theta_{\tilde t} < \pi\,, ~~~  3 < \tan\beta < 60\,, \nn \\
&& 1\TeV < m_A < 10\TeV, ~~~ -10 < \kappa < 10\,.
\end{eqnarray}
Note that even though the decay rate for $\tilde{\sigma}\to\gamma\gamma$ does not depend on the MSSM parameters, ${\rm Br} (\tilde{\sigma}\to\gamma\gamma)$ varies significantly within the range $[0.2,6]\times 10^{-3}$ since the total decay rate changes according to the MSSM parameters.  Our point in the parameter space {\bf P1} has sizable branching ratios for both channels. Below we will also consider two more pessimistic scenarios,
{\bf P2} and {\bf P3}. The point {\bf P2} corresponds to a scenario in which ${\rm Br}(\tilde{\sigma}\to\gamma\gamma)$ gets its minimal value. In this case there is a unique  ${\rm Br}(\tilde{\sigma}\to ZZ )$.  The point {\bf P3} corresponds to ${\rm Br}(\tilde{\sigma}\to Z Z )$ close to its minimum, with the minimum possible ${\rm Br}(\tilde{\sigma}\to\gamma\gamma)$ subject to this constraint\footnote{\baselineskip 3.0ex The branching ratios of $ZZ$ and $\gamma\gamma$ channel are
$0.14$, $~2.5\times 10^{-3}$ respectively for {\bf P1}, $~0.10$, $~0.21\times 10^{-3}$ for {\bf P2}, and $0.0025$, $~1.8\times 10^{-3}$ for {\bf P3}.}. These three benchmark points of the MSSM parameter set are shown in figure \ref{fig:MSSMparam}(a). Note that there is no point in the parameter space where both branching fractions
are negligible. There is a curve, which is roughly a  straight line,  connecting points {\bf P2} and {\bf P3} that forms the boundary of the scatter plot. Moving along this curve one compensates for decreases in one branching ratio with increases in the other. It is clear that points along this curve correspond to  worst-case scenarios for searching for stoponium in these channels: if we can exclude the existence of stoponium for parameter sets along this curve, than this will certainly be true for  the remaining MSSM parameter space. We summarize explicit parameter choices $(\theta_{\tilde{t}}, ~\kappa, ~ m_A, ~ \tan \beta)$ for each benchmark point:
\begin{eqnarray}
{\rm {\bf P1}} &&:~ ( \pi/4,\,-2,~ 2\TeV,~ 10 )\,, \nn \\
{\rm {\bf P2}} &&:~ ( 0.75,~10,~ 2\TeV,~ 10 )\,, \nn \\
{\rm {\bf P3}} &&:~ ( 0.25,\,-9,~ 2\TeV,~ 10 )\,.
\end{eqnarray}
It turns out that each branching ratio is strongly dependent on $\theta_{\tilde{t}}$ and $\kappa$ while the effects of $\tan\beta$ and $m_A$ are minor. For illustration, we show contour plot of ${\rm Br}(\tilde{\sigma}\to\gamma\gamma)$ with respect to $\theta_{\tilde{t}}$ and $\kappa$ in figure \ref{fig:MSSMparam}(b).

We now try to estimate the required luminosity to discover  the stoponium resonance  at current and future LHC runs.
 The required luminosity is evaluated by demanding $5\,\sigma$ significance for the signal events.  We use the following formula for the significance $Z$~\cite{Cowan:2010js}
\begin{equation}
\label{eq:sig}
Z=\sqrt{2\bigg((s+b)\ln\Big(1+\frac{s}{b}\Big) -s\bigg)}\,,
\end{equation}
where $s$ and $b$ represent the number of signal and background events. It is known that this formula is more reliable compared to the commonly used $Z=s/\sqrt{b}$ when the number of background events is small. If $b \gg s$, Eq. (\ref{eq:sig}) reduces to $Z=s/\sqrt{b}$.

We take into account the reconstruction efficiency for photons, $\epsilon_\gamma$, in generating signal and background events for the $\gamma\gamma$ channel. We set $\epsilon_\gamma = 97\%$ as given in Ref. \cite{Aad:2012tfa}. For the $ZZ$ channel, we consider the $ZZ \to 4l (4e, 4\mu, ee\mu\mu)$  final states for reconstructing $ZZ$. We multiply the calculated cross section for $ZZ$  by the branching fractions for decaying to leptons and  the four lepton selection efficiency of $61\%$ as given in Ref. \cite{Collaboration:2012iua}. The generated SM background in the four lepton channels includes the virtual photon contribution.
We expect that including  $Z \to jj $ channels will provide better statistics for the signal yield especially in the high mass region as discussed in Ref. \cite{Aaltonen:2011jn}.

We consider five different stoponium masses ranging between 400 and $800\GeV$ using the MSSM parameter values of the three benchmark points for both 8 and $14\TeV$. This analysis will help determine the better search channel for  the stoponium resonance signal for a given point in MSSM parameter space  and for a given integrated luminosity. The results  are shown in Tables \ref{tab:resonantsignal1} and \ref{tab:resonantsignal2}.
\begin{table}[t]
\centering
\begin{tabular}{ccccccc}
\hline
\hline
\footnotesize $\sqrt{s}=8\TeV$
&~\footnotesize $m_{\tilde{\sigma}}$
&\footnotesize 400 GeV
&\footnotesize 500 GeV
&\footnotesize 600 GeV
&\footnotesize 700 GeV
&\footnotesize 800 GeV\\
\hline
&~\footnotesize \rm{\bf {P1}}
&~\footnotesize 0.62~(171)
&~\footnotesize 0.13~(1197)
&~\footnotesize 0.032~(9654)
&~\footnotesize 0.008~($\ast$)
&~\footnotesize 0.002~($\ast$)\\
\footnotesize $\sigma_{\rm{res}}(pp\to \tilde\sigma \to \gamma\gamma)$
&~\footnotesize \rm{\bf {P2}}
&~\footnotesize 0.018~($\ast$)
&~\footnotesize 0.011~($\ast$)
&~\footnotesize 0.006~($\ast$)
&~\footnotesize 0.002~($\ast$)
&~\footnotesize 0.001~($\ast$)\\

&~\footnotesize \rm{\bf {P3}}
&~\footnotesize 0.55~(212)
&~\footnotesize 0.097~(2186)
&~\footnotesize 0.020~($\ast$)
&~\footnotesize 0.004~($\ast$)
&~\footnotesize 0.001~($\ast$)\\
\cline{2-7}
\footnotesize $\sigma_{\rm{SM}}(pp\to\gamma\gamma)$
&
&\footnotesize 2.1
&\footnotesize 0.70
&\footnotesize 0.34
&\footnotesize 0.14
&\footnotesize 0.06\\
\hline
&~\footnotesize \rm{\bf {P1}}
&~\footnotesize 0.085~(519)
&~\footnotesize 0.034~(1288)
&~\footnotesize 0.015~(3268)
&~\footnotesize 0.006~(8171)
&~\footnotesize 0.003~($\ast$)\\
\footnotesize $\sigma_{\rm{res}}(pp\to \tilde\sigma \to ZZ \to 4l)$
&~\footnotesize \rm{\bf {P2}}
&~\footnotesize 0.032~(3016)
&~\footnotesize 0.025~(2232)
&~\footnotesize 0.016~(2801)
&~\footnotesize 0.008~(5268)
&~\footnotesize 0.004~($\ast$)\\

&~\footnotesize \rm{\bf {P3}}
&~\footnotesize 0.002~($\ast$)
&~\footnotesize $\ast$~($\ast$)
&~\footnotesize $\ast$~($\ast$)
&~\footnotesize $\ast$~($\ast$)
&~\footnotesize $\ast$~($\ast$)\\
\cline{2-7}
\footnotesize $\sigma_{\rm{SM}}(pp\to Z/\gamma^*Z/\gamma^*\to 4l)$
&
&\footnotesize 0.067 &\footnotesize 0.027
&\footnotesize 0.013 &\footnotesize 0.006 &\footnotesize 0.003\\
\hline
\hline
\end{tabular}
\caption{
\baselineskip 3.0ex
The resonant cross section (fb) and the required luminosity (${\rm fb}^{-1}$) for $5\sigma$ discovery in parentheses are shown for each stoponium decay channel at the $8\TeV$ LHC run. The asterisk denotes that the signal cross section is less than $10^{-3} \,{\rm fb}$ or the required luminosity is greater than 10$~{\rm ab}^{-1}$ which means beyond future LHC reach. Several stoponium masses are chosen to be investigated for each benchmark point. For comparison, the integrated cross section within the same invariant mass region of SM background is also shown.
The numbers for cross sections do not include efficiency factors.
}
\label{tab:resonantsignal1}
\end{table}

\begin{table}[t]
\centering
\begin{tabular}{ccccccc}
\hline
\hline
$\footnotesize\sqrt{s}=14\TeV$
&~\footnotesize $m_{\tilde{\sigma}}$
&\footnotesize 400 GeV
&\footnotesize 500 GeV
&\footnotesize 600 GeV
&\footnotesize 700 GeV
&\footnotesize 800 GeV\\
\hline
&~\footnotesize \rm{\bf {P1}}
&~\footnotesize 2.3~(28)
&~\footnotesize 0.55~(173)
&~\footnotesize 0.15~(956)
&~\footnotesize 0.043~(6650)
&~\footnotesize 0.014~($\ast$)\\
~\footnotesize $\sigma_{\rm{res}}(pp\to \tilde\sigma \to \gamma\gamma)$
&~\footnotesize \rm{\bf {P2}}
&~\footnotesize 0.067~($\ast$)
&~\footnotesize 0.047~($\ast$)
&~\footnotesize 0.026~($\ast$)
&~\footnotesize 0.012~($\ast$)
&~\footnotesize 0.005~($\ast$)\\

&~\footnotesize \rm{\bf {P3}}
&~\footnotesize 2.0~(34)
&~\footnotesize 0.40~(312)
&~\footnotesize 0.088~(2534)
&~\footnotesize 0.022~($\ast$)
&~\footnotesize 0.006~($\ast$)\\
\cline{2-7}
\footnotesize$\sigma_{\rm{SM}}(pp\to\gamma\gamma)$
    &
&\footnotesize 4.4
&\footnotesize 1.7
&\footnotesize 0.67
&\footnotesize 0.41
&\footnotesize 0.21\\

\hline

&~\footnotesize \rm{\bf {P1}}
&~\footnotesize 0.31~(104)
&~\footnotesize 0.14 ~(223)
&~\footnotesize 0.066~(489)
&~\footnotesize 0.032~(1076)
&~\footnotesize 0.017~(2261)\\
~\footnotesize $\sigma_{\rm{res}}(pp\to \tilde\sigma \to ZZ \to 4l)$
&~\footnotesize \rm{\bf {P2}}
&~\footnotesize 0.12~(572)
&~\footnotesize 0.10~(378)
&~\footnotesize 0.073~(422)
&~\footnotesize 0.041~(707)
&~\footnotesize 0.022~(1413)\\

&~\footnotesize \rm{\bf {P3}}
&~\footnotesize 0.007~($\ast$)
&~\footnotesize 0.002~($\ast$)
&~\footnotesize $\ast$~($\ast$)
&~\footnotesize $\ast$~($\ast$)
&~\footnotesize $\ast$~($\ast$)\\
\cline{2-7}
\footnotesize$\sigma_{\rm{SM}}(pp\to Z/\gamma^*Z/\gamma^*\to 4l)$
&
&\footnotesize 0.16
&\footnotesize 0.070
&\footnotesize 0.035
&\footnotesize 0.019
&\footnotesize 0.010\\
\hline
\hline
\end{tabular}
\caption{
\baselineskip 3.0ex
The resonant cross section (fb) and the required luminosity (${\rm fb}^{-1}$) for $5\sigma$ discovery in parentheses are shown for each stoponium decay channel at the $14\TeV$ LHC run. The asterisk denotes that the signal cross section is less than $10^{-3} \,{\rm fb}$ or the required luminosity is greater than 10$~{\rm ab}^{-1}$ which means beyond future LHC reach. Several stoponium masses are chosen to be investigated for each benchmark point. For comparison, the integrated cross section within the same invariant mass region of SM background is also shown. The numbers for cross sections do not include efficiency factors.
}
\label{tab:resonantsignal2}
\end{table}

For the $8\TeV$ run, the required luminosity rapidly grows as the stoponium mass increases in any of the parameter sets. It easily goes beyond the current and future LHC reach. We see that it is hopeless to observe the resonance signal for stoponium masses between 400 and $800\GeV$ in any of the  benchmark parameter sets with the current accumulated LHC luminosity, $23\,{\rm fb}^{-1}$.

On the other hand, for the $14\TeV$ run, we are able to explore the stoponium mass up to 500$\GeV$ in the first round of a future LHC run with $400~{\rm fb}^{-1}$ of accumulated data. As we expect, for the parameter sets {\bf P1} and {\bf P2}, the $ZZ$ channel is the most promising for discovering stoponium, while in the case of {\bf P3} the $\gamma\gamma$ channel is better  than the $ZZ$ channel.
In order, for example, to see a 500$\GeV$ stoponium resonance signal we need at least 378$~{\rm fb}^{-1}$ using both $\gamma\gamma$ and $ZZ$ channels regardless of the MSSM parameter sets. For a heavier stoponium mass, one needs the high-luminosity LHC run with upgraded instantaneous luminosity. We note if the expected stoponium resonance signal is not observed in the $\gamma\gamma$ and/or $ZZ$ channels we can exclude some of the MSSM parameter space in the $\theta_{\tilde t}$-$\kappa$ plane.

\section{Conclusion}
In this work we have studied the production of stoponium in  $pp$  collisions at the LHC. Our analysis focused on the MSSM scenario with light stops in the $200$-$400$ GeV mass range, which are able to evade existing searches that look for stops that decay to neutralinos and top quarks. We used effective field theory  to obtain a factorized form of the cross section.
This allowed us to resum large threshold logarithms using RG equation methods up to NNLL accuracy. We verified explicitly that when NLO+NNLL results are considered, theoretical uncertainties
are considerably reduced which highly improves the phenomenological predictions for the LHC.
We included the enhanced Coulomb interactions responsible for creating the bound state of stops, stoponium,
by including the NLO strong Coulomb Green`s function. We provided formulas for both total and differential cross sections.

On the phenomenological side, we considered the decays of  stoponium to $\gamma\gamma$ and $ZZ$ as promising channels for searching for the stoponium resonance in the mass range 400 to $800 \GeV$.
After investigating MSSM parameter dependence, we found that $\gamma\gamma$ and $ZZ$ channels sensitively depend on $\theta_{\tilde t}$ and $\kappa$ while the effects of other MSSM parameters are negligible. Therefore, one can impose constraints on $\theta_{\tilde t}$ and $\kappa$ if the stoponium resonance signal is not observed.
Our results indicated that one cannot exclude any mass value in this mass range with the currently accumulated LHC data at $8\TeV$.  On the other hand, for the first round of future LHC runs at $14\TeV$ with $400\,{\rm fb}^{-1}$ integrated luminosity, it should be possible to find stoponium if its mass is less than  $500\GeV$ via either the $\gamma\gamma$ or $ZZ$ channels. We stress that this result does not depend on any particular choice of MSSM parameters. In this regard, searching  in $\gamma \gamma $ and $ZZ$ for
the stoponium resonance  will serve as a complementary method for probing 
light stop scenarios
in  future LHC runs.

\begin{acknowledgements}
We would like to thank Stephen Martin for pointing out some subtle issues regarding the phenomenological part of our work.
C.~Kim was supported by the Basic Science Research Program through NRF funded by MSIP (Grant No. 2012R1A1A1003015).
C.~Kim thanks KIAS for its hospitality during a visit to complete this work.
A. Idilbi is supported by the U.S Department of Energy under Grant No. DE-SC0008745.
T. Mehen is supported in part by the U.S. Department of Energy, Office of Nuclear Physics,  under Contract No. DE-FG02-05ER41368.
T. Mehen thanks ECT* for its hospitality  during which part of this work was performed.
Y.W.~Yoon thanks the KIAS Center for Advanced Computation for providing computing resources.
Y.W.~Yoon thanks the L/EFT theory group of Duke University for its hospitality during a visit where part of this work was done.

\end{acknowledgements}

\appendix
\section{Stoponium decay rate}
Here we present the relevant formulas of stoponium decay rates at LO in $\alpha_s$. As we mentioned in Sec.~V,
we neglect contributions from another heavier stop, gluino and a heavy Higgs by assuming that they are much heavier than a light stop.
The stoponium decay rate can be related with the cross section of stop pair annihilation through $\Gamma(\tilde\sigma \to AB) = {\mathrm v}\sigma({\tilde t}{\tilde t}^* \to AB) |\psi(0)|^2$, where $\mathrm v$ is the relative velocity between stops and $\psi(0)$ is a stop bound state wave function at the origin. Then the decay rate has following conventional form
\begin{equation}
\Gamma(\tilde\sigma \to AB) = \frac{N_c}{8\pi (1+\delta_{AB})} \frac{|\psi(0)|^2}{m_{\tilde \sigma}^2} \lambda^{\frac{1}{2}}\bigg(1,\frac{m_A^2}{m_{\tilde \sigma}^2},\frac{m_B^2}{m_{\tilde \sigma}^2}\bigg)|{\bar {\cal M}}(\tilde\sigma \to AB)|^2\,,
\end{equation}
where the triangle function $\lambda(x,y,z)$ is given as $\lambda(x,y,z) = x^2+y^2+z^2-2xy-2yz-2zx$.

In the small velocity limit, the matrix elements squared of each decay channel  are
{\allowdisplaybreaks
\begin{eqnarray}
|{\bar {\cal M}}(\tilde\sigma \to gg)|^2 &=& 2 g_s^4 \frac{(N_c^2-1)}{N_c^2} \,, \\
|{\bar {\cal M}}(\tilde\sigma \to \gamma\gamma)|^2 &=& 8 Q_{\tilde t}^4 e^4 \,, \\
|{\bar {\cal M}}(\tilde\sigma \to ZZ)|^2 &=& 2\left| g_{{\tilde t}{\tilde t}ZZ} - \frac{ \lambda_{{\tilde t}{\tilde t}h} \lambda_{hZZ}}{4m_{\tilde t}^2-m_h^2} \right|^2 \nn \\
&& + \left| \bigg(1-\frac{2m_{\tilde t}^2}{m_Z^2}\bigg) \bigg( g_{{\tilde t}{\tilde t}ZZ} - \frac{ \lambda_{{\tilde t}{\tilde t}h} \lambda_{hZZ}}{4m_{\tilde t}^2-m_h^2} \bigg) + 8 g_{{\tilde t}{\tilde t}Z}^2 \frac{m_{\tilde t}^2(m_{\tilde t}^2 - m_Z^2)}{m_Z^2(2m_{\tilde t}^2 - m_Z^2)}\right|^2 \,, \\
|{\bar {\cal M}}(\tilde\sigma \to WW)|^2 &=& 2\left| g_{{\tilde t}{\tilde t}WW} - \frac{ \lambda_{{\tilde t}{\tilde t}h} \lambda_{hWW}}{4m_{\tilde t}^2-m_h^2} \right|^2 + \left| \bigg(1-\frac{2m_{\tilde t}^2}{m_W^2}\bigg) \bigg( g_{{\tilde t}{\tilde t}WW} - \frac{ \lambda_{{\tilde t}{\tilde t}h} \lambda_{hWW}}{4m_{\tilde t}^2-m_h^2} \bigg) \right|^2 \,, \\
|{\bar {\cal M}}(\tilde\sigma \to Z\gamma)|^2 &=& 3g_{{\tilde t}{\tilde t}Z\gamma}^2 - 4 Q_{\tilde t}^2 e^2 g_{{\tilde t}{\tilde t}Z}^2 \,, \\
|{\bar {\cal M}}(\tilde\sigma \to hh)|^2 &=&  \left| \lambda_{{\tilde t}{\tilde t} hh} - \frac{\lambda_{{\tilde t}{\tilde t}h} \lambda_{hhh}}{4m_{\tilde t}^2-m_h^2}+\frac{2\lambda_{{\tilde t}{\tilde t}h}^2}{2m_{\tilde t}^2-m_h^2}\right|^2 \,, \\
|{\bar {\cal M}}(\tilde\sigma \to b{\bar b})|^2 &=& 8N_c\lambda_{{\tilde t }{\tilde t}h}^2\lambda_{hbb}^2 \frac{(m_{\tilde t}^2 - m_b^2)}{(4m_{\tilde t}^2-m_h^2)^2} \,, \\
|{\bar {\cal M}}(\tilde\sigma \to t{\bar t})|^2 &=& 8N_c\lambda_{{\tilde t }{\tilde t}h}^2\lambda_{htt}^2 \frac{(m_{\tilde t}^2 - m_t^2)}{(4m_{\tilde t}^2-m_h^2)^2} \,.
\end{eqnarray}
}
Here, $Q_{\tilde t}$ is the electric charge of the stop. The MSSM coupling constants are defined as
{\allowdisplaybreaks
\begin{eqnarray}
g_{{\tilde t}{\tilde t}Z} &=& \frac{e}{2 c_W s_W}\Big(|c_{\tilde t}|^2 - \frac{4}{3} s_W^2 \Big)\,,\\
g_{{\tilde t}{\tilde t}ZZ} &=& \frac{2e^2}{3 c_W^2 }\bigg(\frac{(3-8s_W^2)}{4s_W^2}|c_{\tilde t}|^2 + \frac{4}{3} s_W^2 \bigg)\,,\\
g_{{\tilde t}{\tilde t}WW} &=& \frac{e^2}{2 s_W^2} |c_{\tilde t}|^2 \,, \\
g_{{\tilde t}{\tilde t}Z\gamma} &=& \frac{Q_{\tilde t} e^2}{ c_W s_W} \Big( |c_{\tilde t}|^2 -\frac{4}{3}s_W^2\Big) \,, \\
\lambda_{hZZ} &=& \frac{e^2}{2 c_W^2 s_W^2} \sin(\beta-\alpha) v\,, \\
\lambda_{hWW} &=& \frac{e^2}{2 s_W^2} \sin(\beta-\alpha) v\,, \\
\lambda_{hbb} &=& \frac{\sin\alpha}{\cos\beta} \frac{m_b}{v} \,, \\
\lambda_{htt} &=& -\frac{\cos\alpha}{\sin\beta} \frac{m_t}{v} \,, \\
\lambda_{hhh} &=& -\frac{3e^2}{4c_W^2 s_W^2} \sin(\beta+\alpha)\cos(2\alpha) v\,, \\
\lambda_{{\tilde t}{\tilde t}h} &=& \frac{e^2}{3c_W^2}\sin(\alpha+\beta)\bigg(1+\frac{(3-8s_W^2)}{4s_W^2}|\,c_{\tilde t}|^2\bigg)v - \frac{2\cos\alpha}{\sin\beta} \frac{m_t^2}{v} -  \frac{2{\rm Re} [c_{\tilde t} s_{\tilde t}\kappa m_W]}{\sin\beta} \frac{m_t}{v}\,, \\
\lambda_{{\tilde t}{\tilde t}hh} &=& \frac{e^2}{3c_W^2 }\cos2\alpha\bigg(1+\frac{(3-8s_W^2)}{4s_W^2}|\,c_{\tilde t}|^2\bigg)- \frac{2\cos^2\alpha}{\sin^2\beta} \frac{m_t^2}{v^2}\,,
\end{eqnarray}
}
where $s_W, c_W$ are the sine and cosine of the weak mixing angle, and $s_{\tilde t}$ and $c_{\tilde t}$ are the sine and cosine of stop
mixing angle $\theta_{\tilde t}$. $v$ is the Higgs VEV, fixed by $v= (\sqrt{2} G_F)^{-1/2} = 246\GeV$. We refer to Ref. \cite{Rosiek:1995kg}
to obtain the Feynman rules for the MSSM.

\section{Anomalous dimensions}
In this section we summarize all the formulas for the anomalous dimensions that are necessary to perform resummation  up to NNLL.
First of all, the function $\beta(\alpha_s)=d\alpha_s/d\ln \mu$ is expanded in $\alpha_s$ as
\begin{eqnarray}
\beta(\alpha_s) = -2\alpha_s  \sum_{k=0}^\infty \beta_k \Big( \frac{\alpha_s}{4\pi} \Big)^{k+1}\,,
\end{eqnarray}
where we have
\begin{eqnarray}
\beta_0 &=& \frac{11}{3}C_A - \frac{4}{3} T_F n_f \,, \nn \\
\beta_1 &=& \frac{34}{3}C_A^2-\frac{20}{3} C_A T_F n_f - 4 C_F T_F n_f\,, \nn \\
\beta_2 &=& \frac{2857}{54} C_A^3 + \Big( 2C_F^2-\frac{205}{9} C_F C_A - \frac{1415}{27} C_A^2\Big)T_F n_f + \Big( \frac{44}{9}C_F + \frac{158}{27} C_A \Big) T_F^2 n_f^2\,.
\end{eqnarray}

The convention for the expansion of anomalous dimension $\gamma^A$ in $\alpha_s$ is
\begin{eqnarray}
\gamma^A &=& \sum_{k=0} \gamma_k^A\Big(\frac{\alpha_s}{4\pi} \Big)^{k+1}\,.
\end{eqnarray}
The anomalous dimension $\gamma^S$ for hard scattering of Higgs boson production is equivalent to that of stoponium production since the effective
theory calculations are the same at the hard matching scale (whether it is the mass of the Higgs boson or the stop pair). The coefficients of $\gamma^S$ are   \cite{Idilbi:2005ni,Idilbi:2005er}
\begin{eqnarray}
\gamma_0^S &=& 0 \,, \nn \\
\gamma_1^S &=& C_A^2\Big( -\frac{160}{27} + \frac{11\pi^2}{9} + 4\zeta_3\Big) + C_A T_F n_f \Big(-\frac{208}{27}-\frac{4\pi^2}{9}\Big) -8C_F T_F n_f \,,
\end{eqnarray}
up to two-loop order. The anomalous dimension for the soft function $\gamma^W$ is obtained with the aid of the relation: $\gamma^W = \frac{\beta(\alpha_s)}{\alpha_s}+ 2\gamma^B + \gamma^S$, where $2\gamma^B$ is the coefficient of the $\delta(1-x)$ term in the Altarelli-Parisi splitting function $P_{gg}(x)$ which reads \cite{Vogt:2004mw}
\begin{eqnarray}
\gamma_0^B &=& \beta_0\,, \nn \\
\gamma_1^B &=& 4 C_A^2\Big(\frac{8}{3} + 3\zeta_3\Big) - \frac{16}{3} C_A T_F n_f - 4 C_F T_F n_f \,.
\end{eqnarray}

The solution for the Sudakov exponent $S_\Gamma(\mu_1,\mu_2)$ is expressed as \cite{Becher:2007ty}
\begin{eqnarray}
S_\Gamma(\mu_1,\mu_2) &=& \frac{\Gamma_{C,0}^A}{4\beta_0^2}\Bigg[
\frac{4\pi}{\alpha_s(\mu_1)} \Big( 1 - \frac{1}{r} - \ln r \Big) + \Big( \frac{\Gamma_{C,1}^A}{\Gamma_{C,0}^A}-\frac{\beta_1}{\beta_0} \Big) (1-r+\ln r) + \frac{\beta_1}{2\beta_0} \ln^2 r \nn \\
&& ~~ + \frac{\alpha_s(\mu_1)}{4\pi} \Bigg\{ \Big(\frac{\beta_1 \Gamma_{C,1}^A}{\beta_0 \Gamma_{C,0}^A} - \frac{\beta_2}{\beta_0} \Big) (1-r+r\ln r) + \Big(\frac{\beta_1^2}{\beta_0^2} - \frac{\beta_2}{\beta_0}\Big)(1-r)\ln r \nn \\
&& ~~ - \Big(\frac{\beta_1^2}{\beta_0^2} - \frac{\beta_2}{\beta_0} - \frac{\beta_1 \Gamma_{C,1}^A}{\beta_0 \Gamma_{C,0}^A} + \frac{\Gamma_{C,2}^A}{\Gamma_{C,0}^A} \Big) \frac{(1-r)^2}{2} \Bigg\} +\ldots
\Bigg]\,,
\end{eqnarray}
where $r = \alpha_s(\mu_2)/\alpha_s(\mu_1)$. The coefficients of the expansion (in $\alpha_s$) of the cusp anomalous dimension $\Gamma_C^A$ for the Wilson
loop in the adjoint representation (up to third order in $\alpha_s$) read \cite{Korchemskaya:1992je,Moch:2004pa}
\begin{eqnarray}
\Gamma_{C,0}^A &=& 4 C_A \,,\nn \\
\Gamma_{C,1}^A &=& 4 C_A \Bigg[ \Big( \frac{67}{9} - \frac{\pi^2}{3} \Big)C_A - \frac{20}{9} T_F n_f \Bigg] \,,\nn \\
\Gamma_{C,2}^A &=& 4 C_A \Bigg[ C_A^2 \Big( \frac{245}{6} - \frac{134 \pi^2}{27} + \frac{11 \pi^4}{45} + \frac{22}{3}\zeta_3 \Big) + C_A T_F n_f \Big( -\frac{418}{27} + \frac{40 \pi^2}{27} - \frac{56}{3}\zeta_3 \Big) \nn \\
&& ~~~~ + C_F T_F n_f \Big(-\frac{55}{3} + 16 \zeta_3\Big) - \frac{16}{27} T_F^2 n_f^2\Bigg]\,.
\end{eqnarray}
The solution for the exponent $a_\Gamma (\mu_1, \mu_2 )$ is expressed as
\begin{eqnarray}
a_\Gamma (\mu_1, \mu_2 ) &=& \frac{\Gamma_{C,0}^A}{2\beta_0} \Bigg[ \ln \frac{\alpha_s(\mu_2)}{\alpha_s(\mu_1)} + \Big( \frac{\Gamma_{C,1}^A}{\Gamma_{C,0}^A} - \frac{\beta_1}{\beta_0}\Big) \frac{\alpha_s(\mu_2)-\alpha_s(\mu_1)}{4\pi} \nn \\
&& ~~~~~~ +\Bigg( \frac{\Gamma_{C,2}^A}{\Gamma_{C,0}^A} - \frac{\beta_2}{\beta_0} - \frac{\beta_1}{\beta_0}\Big(\frac{\Gamma_{C,1}^A}{\Gamma_{C,0}^A}-\frac{\beta_1}{\beta_0}\Big)\Bigg) \frac{\alpha_s^2(\mu_2)-\alpha_s^2(\mu_1)}{32\pi^2}+\ldots
\Bigg]\,.
\end{eqnarray}
The exponent $a_{\gamma^W}(\mu_1,\mu_2)$  can also  be expressed in a similar way.

\section{Coulomb Green's Function}

In this section, we present the formula of the Greens's function up to NLO in $\alpha_s$ that is used in phenomenological analysis of this work. Especially, we explicitly show the analytic continuation of the Green's function at NLO which is essential for numerical implementation of it when the bound state resonance is narrow.

The Green's function of the Schr\"odinger equation with Coulomb potential up to NLO between heavy colored particles was calculated in Ref. \cite{Beneke:1999qg}. We refer to Refs. \cite{Kiyo:2008bv,Beneke:2011mq} for the explicit formula of the Greens's function at the origin $G_{\bf 1}(0,0,E+i\Gamma)$ for the color-singlet stoponium bound state which is described by
\begin{equation}
G_{\bf 1}(0,0,E+i\Gamma) = C_F m_{\tilde t}^2 \frac{\alpha_s(\mu_C)}{4\pi}\bigg( G^{(0)}_{\bf 1}(E+i\Gamma) + \frac{\alpha_s(\mu_C)}{4\pi} G^{(1)}_{\bf 1}(E+i\Gamma)\bigg).
\end{equation}
The LO and NLO Green's functions $G^{(0)}_{\bf 1}$ and $G^{(1)}_{\bf 1}$ read
\begin{eqnarray}\label{GFLO}
G^{(0)}_{\bf 1}(E+i\Gamma) &=& -\frac{1}{2\lambda} + L - \psi^{(0)}(1-\lambda)\,, \\
G^{(1)}_{\bf 1}(E+i\Gamma) &=& \beta_0 \Big[ L^2 - 2L \big( \psi^{(0)}(1-\lambda) - \lambda\, \psi^{(1)}(1-\lambda) \big) + \lambda\,\psi^{(2)}(1-\lambda)
 \nn \\&&~~
+ \big(\psi^{(0)}(1-\lambda)\big)^2
 -3\psi^{(1)}(1-\lambda)-2\lambda\psi^{(0)}(1-\lambda)\psi^{(1)}(1-\lambda)
 \nn \\ &&~~
 +4\,{}_4F_3(1,1,1,1\,;2,2,1-\lambda\,;1)\Big]
 \nn \\ &&
 +a_1 \Big[ L - \psi^{(0)}(1-\lambda) + \lambda \psi^{(1)}(1-\lambda) \Big]\,,
\end{eqnarray}
where $a_1 = \frac{31}{9} C_A - \frac{20}{9} T_F n_f $. $L$ and $\lambda$ are defined by
\begin{eqnarray}
L= \ln \bigg( \frac{i\mu_C}{2m_{\tilde t} \bar v} \bigg), ~~ \lambda = \frac{ i C_F \alpha_s(\mu_C)}{2 \bar v}, ~~ \bar v = \sqrt{\frac{E+i\Gamma}{m_{\tilde t}}}\,.
\end{eqnarray}
The functions $\psi^{(n)}(z) $ are defined by
\begin{equation}
\psi^{(0)}(z) = \gamma_E + \frac{d}{dz}\ln \Gamma(z), ~~ \psi^{(n)}(z) = \frac{d^n}{dz^n} \psi^{(0)}(z)\,.
\end{equation}
The appropriate Coulomb scale is estimated by $\mu_C = m_{\tilde t} C_F \alpha_s(\mu_C)$.

It is nontrivial to evaluate the hypergeometric function in the NLO Green's function. The series expansion of the hypergeometric function with unit argument ${}_PF_{P-1}(a_1,\ldots,a_{P}\,;b_1,\ldots,b_{P-1}\,;1)$ is convergent if
\begin{equation}
\label{eq:Sval}
S \equiv {\rm Re} \sum_{j=1}^{P-1}b_j - {\rm Re}\sum_{j=1}^P a_j > 0\,.
\end{equation}
Therefore, the hypergeometric function ${}_4F_3(1,1,1,1\,;2,2,1-\lambda\,;1)$ is well defined in $\rm Re \lambda < 1$.
Since we encounter $\rm Re \lambda \geq 1$ above the resonance peak, we need to analytically continuate the hypergeometric function into the complex plane in which $\rm Re \lambda \geq 1$.

In general, the hypergeometric function with unit argument can be analytically continuated by using the relation \cite{Huber:2005yg}
\begin{eqnarray}
&&\Big( \sum_{j=1}^{P-1}b_j - \sum_{j=1}^P a_j\Big)  {}_PF_{P-1}(a_1,\ldots,a_{P}\,;b_1,\ldots,b_{P-1}\,;1) = \nn \\
&&~~~ \sum_{j=1}^{P-1} \frac{\prod_{k=1}^P (b_j - a_k)}{\prod_{l=1,l\ne j}^{P-1}(b_j - b_l)}\, \frac{1}{b_j}  \,{}_PF_{P-1}(a_1,\ldots,a_{P}\,;b_1,\ldots,b_{j-1},b_j +1,b_{j+1}, \ldots,b_{P-1}\,;1).
\end{eqnarray}
We note that the $S$ value of Eq. (\ref{eq:Sval}) in the right-hand side is increased by 1. Thus, by repeatedly applying this relation one can represent any hypergeometric function with unit argument ${}_PF_{P-1}(a_1,\ldots,a_{P}\,;b_1,\ldots,b_{P-1}\,;1)$ in terms of convergent hypergeometric functions.

On the other hand, the method described in Ref. \cite{Beneke:2011mq} is useful for investigating the asymptotic behavior of the Green's function near rhe resonance region. The explicit expansion of ${}_4F_3(1,1,1,1\,;2,2,1-\lambda\,;1)$ is given by
\begin{equation}
F_{43} \equiv {}_4F_3(1,1,1,1\,;2,2,1-\lambda\,;1) = \sum_{i=0}^\infty \frac{\Gamma(i+1)^3\Gamma(1-\lambda)}{\Gamma(i+2)^2 \Gamma(1-\lambda+i)}\,.
\end{equation}
This infinite sum can be handled by Ref. \cite{Bierenbaum:2008yu} and be represented in terms of harmonic sums \cite{Vermaseren:1998uu,Blumlein:1998if},
\begin{equation}
F_{43} = \zeta(2)- S_2(-\lambda) - \lambda \Big[ \zeta(3) + S_3(-\lambda) - S_1(-\lambda)\Big(\zeta(2)-S_2(-\lambda)\Big) - S_{2,1}(-\lambda) \Big]\,.
\end{equation}
The harmonic sums $S_n(-\lambda)$ are related with $\psi^{(n-1)}$ functions as follows
\begin{eqnarray}
S_1(-\lambda) &=& \psi^{(0)}(1-\lambda)\,, \\
S_n(-\lambda) &=& \frac{(-1)^{n-1}}{\Gamma(n)} \psi^{(n-1)}(1-\lambda) + \zeta(n),  ~~ (n \ge 2)\,.
\end{eqnarray}
The nested harmonic sum $S_{2,1}(-\lambda)$ can be represented in integral form \cite{Blumlein:2009ta},
\begin{eqnarray}
S_{2,1}(-\lambda) = -\int_0^1 dx \bigg(\frac{x^{-\lambda} -1}{x-1}\bigg)\,{\rm Li}_2(x) + \zeta(2)S_1(-\lambda)\,.
\end{eqnarray}
This integral form is well defined for ${\rm Re} \lambda < 2$. The analytic continuation into ${\rm Re} \lambda \geq 2$ is straightforward:
\begin{eqnarray}
S_{2,1}(-\lambda) &=& -\int_0^1 dx \bigg(\frac{x^{-\lambda} -1}{x-1}\bigg)\,\bigg({\rm Li}_2(x)-\sum_{j=1}^{m-1}
 \frac{x^j}{j^2}\bigg)
 \nn \\ &&~~~
- \sum_{j=1}^{m-1}\frac{S_1(j-\lambda)}{j^2} + S_{2,1}(m-1) + \zeta(2)S_1(-\lambda)\,,
\end{eqnarray}
where positive integer $m$ is chosen such that $ {\rm Re} \lambda < m+1$. In all, we finally obtain an expression of an analytically continuated NLO Green's function which can be evaluated in any complex value of $\lambda\,$:
\begin{eqnarray}\label{GFNLO}
G^{(1)}_{\bf 1}(E+i\Gamma) &=& \beta_0 \Bigg[ L^2 - 2L \big( \psi^{(0)}(1-\lambda) - \lambda\, \psi^{(1)}(1-\lambda) \big) - \lambda\,\psi^{(2)}(1-\lambda)
 \nn \\&&~~
+ \big(\psi^{(0)}(1-\lambda)\big)^2
 +\psi^{(1)}(1-\lambda) + 2\lambda\psi^{(0)}(1-\lambda)\big(2\zeta(2)+ \psi^{(1)}(1-\lambda)\big)
 \nn \\ &&~~
 -8\lambda\zeta(3) -\zeta(2)
 +4\lambda\Bigg(
  S_{2,1}(m-1)
 -\int_0^1 dx \bigg(\frac{x^{-\lambda} -1}{x-1}\bigg)\,\bigg({\rm Li}_2(x)-\sum_{j=1}^{m-1}
 \frac{x^j}{j^2}\bigg)
 \nn \\ &&~~
 - \sum_{j=1}^{m-1}\frac{\psi^{(0)}(j+1-\lambda)}{j^2}\Bigg)\Bigg]
 +a_1 \Big[ L - \psi^{(0)}(1-\lambda) + \lambda \psi^{(1)}(1-\lambda) \Big]\,.
\end{eqnarray}
We note that the resonance peaks arise when $\rm Re{\lambda}$ becomes positive integer values, that is ${\rm Re} \lambda = 1, 2, ...$ for narrow resonance states. We use this formula for the Green's function at NLO throughout the numerical analysis of this work.

\end{document}